\documentclass[usenatbib]{mn2e}
\usepackage{psfig}
\usepackage[dvips]{graphicx}

\usepackage{amsmath}

\newcommand{\LCDM}{$\Lambda$CDM}
\newcommand{\msun}{\mbox{${\rm M}_{\odot}$}}

\newcommand{\mum}{\mbox{$\mu$m}}

\newcommand{\plottwo}[2]
           {\centering \leavevmode \psfig{file=#1,width=\columnwidth,clip=}
                            \hfill \psfig{file=#2,width=\columnwidth,clip=}}

\def\lesssim{\lower.5ex\hbox{$\; \buildrel < \over \sim \;$}}
\def\gtrsim{\lower.5ex\hbox{$\; \buildrel > \over \sim \;$}}
\begin{document}

\title[Galaxy Properties from the UV to the FIR]{Galaxy Properties
  from the Ultra-violet to the Far-Infrared: \LCDM\ models
  confront observations}

\author[R. S. Somerville, R. C. Gilmore, J.R. Primack, A. Dominguez] {
%\parbox[t]{\textwidth}{ 
Rachel S. Somerville$^{1,2}$\thanks{E-mail: somerville@stsci.edu}, Rudy C. Gilmore$^{3,4}$, Joel R. Primack$^{3}$, 
Alberto Dom\'{i}nguez$^{5,6,7}$\\
%\vspace*{6pt} \\
$^1$ Space Telescope Science Institute, 3700 San Martin Dr., Baltimore, MD 21218\\
$^2$ Department of Physics and Astronomy, Johns Hopkins University, Baltimore, MD 21218\\
$^3$ Department of Physics, University of California, Santa Cruz, CA 95064\\
$^4$ Scuola Internazionale Superiore di Studi Avanzati (SISSA), Via Bonomea 265, 34136, Trieste, Italy\\
$^5$ Visiting researcher at the Santa Cruz Institute for Particle Physics (SCIPP), University of California, Santa Cruz, CA 95064, USA\\
$^6$ Instituto de Astrof\'{i}sica de Andaluc\'{i}a, CSIC, Apdo. Correos 3004, E-18080 Granada, Spain\\
$^7$ Departamento de F\'{i}sica At\'{o}mica, Molecular y Nuclear, Universidad de Sevilla, Apdo. Correos 1065, E-41080 Sevilla, Spain
}

\maketitle

\begin{abstract}

We combine a semi-analytic model of galaxy formation with simple
analytic recipes describing the absorption and re-emission of
starlight by dust in the interstellar medium of galaxies. We use the
resulting models to predict galaxy counts and luminosity functions
from the far-ultraviolet to the sub-mm, from redshift five to the
present, and compare with an extensive compilation of observations. We
find that in order to reproduce the rest-UV and optical luminosity
functions at high redshift, we must assume an evolving normalization
in the dust-to-metal ratio, implying that galaxies of a given
bolometric luminosity (or metal column density) must be less
extinguished than their local counterparts. In our best-fit model, we
find remarkably good agreement with observations from rest $\sim 1500$
\AA\ to $\sim 250\, \mu$m. At longer wavelengths, most dramatically in
the sub-mm, our models underpredict the number of bright galaxies by a
large factor. The models reproduce the observed total IR luminosity
function fairly well.  We show the results of varying several
ingredients of the models, including various aspects of the dust
attenuation recipe, the dust emission templates, and the cosmology. We
use our models to predict the integrated Extragalactic Background
Light (EBL), and compare with an observationally-motivated EBL model
and with other available observational constraints.
\end{abstract}


%=======================
% 1
\section{Introduction}
\label{sec:intro}
%=======================

New observational facilities have greatly extended the range of the
electromagnetic spectrum over which emission from galaxies can be
measured, while simultaneously expanding the range of cosmic history
that can be probed. For example, in recent years, survey observations
over significant fractions of the sky have been carried out in the Far
and Near Ultra-violet by GALEX (Galaxy Evolution Explorer), in the
optical by the Sloan Digital Sky Survey (SDSS), in the Near-infrared
(IR) by 2MASS (Two Micron All Sky Survey), in the Near-IR and mid-IR
by the Spitzer Space Telescope, and most recently, in the far-IR by
the Herschel Telescope. In addition, multiple $~\sim 0.5-1$ sq. degree
sized fields have now been deeply imaged in the X-ray with Chandra, in
the optical with the Advanced Camera for Surveys (ACS) on the Hubble
Space Telescope (HST) as well as ground-based facilities, and in the
near-IR with UKIRT and Spitzer, allowing large samples of high
redshift galaxies to be identified and studied. The recently installed
Wide Field Camera 3 (WFC3) on HST is in the process of observing many
of these fields at high resolution in the Near IR. Crucial to the
extraction of physical quantities and scientific insight from these
deep surveys has also been the availability of accurate redshift
information for large numbers of galaxies from multi-wavelength
medium-band surveys (COMBO-17, COSMOS, MUSYC, NEWFIRM) and
multi-object spectroscopy (DEEP, VIMOS).

However, perhaps some of the most surprising and poorly understood
observational results of the past two decades have come from
long-wavelength observations in the mid- to far-IR. The IRAS satellite
revealed that $\sim30$\% of the bolometric luminosity of nearby
galaxies, mostly normal spirals, is reprocessed by dust in the IR
\citep{soifer91}, and discovered a population of heavily obscured
luminous and ultra-luminous infrared galaxies (LIRGs and ULIRGS;
\citealt{sanders:96}). IRAS already provided hints of the very strong
evolution of this IR-bright population, the number density of which seems to
have been enormously larger in the past. This was confirmed and
quantified first by ISO \citep[e.g.][]{elbaz:99,elbaz02},
%(Elbaz et al.  1999, 2002; Gruppioni et al. 2002; Dole et al. 2001, 
and then by SCUBA \citep{smail:97,hughes:98,chapman:05}
%(Smail et al. 1997, 2002; Hughes et al. 1998; Chapman et al. 2005) 
and Spitzer \citep[e.g.][]{lefloch05,babbedge06}. 
%(e.g. Le Floch et al. 2005; Babbedge et al. 2006). 
The physical interpretation of these high redshift LIRGS and ULIRGS,
however, has until recently been hampered by the fact that in many
cases observations existed only in the mid-IR or sub-mm, relatively
far from the $\sim 100\, \mu$m peak of the dust emission (e.g. 15
$\mu$m in the case of ISO, 24 $\mu$m for Spitzer, and 450 and 850
$\mu$m in the case of SCUBA). This situation should improve greatly in
the next few years, as observations bracketing the 100 $\mu$m peak are
taken by the PACS (57 to 210 $\mu$m) and SPIRE (250, 350, 500 $\mu$m)
instruments on the recently launched Herschel telescope.

This rainbow of observations presents a challenge to theoretical
models of galaxy formation. To date, most studies have focussed on
making predictions for rest-optical or intrinsic (e.g. stellar mass,
star formation rate) properties of galaxies, mainly because of the
difficulty of modeling the absorption and emission of light by dust
in the interstellar medium of galaxies. However, observational
estimates of intrinsic physical properties from multi-wavelength
photometry suffer from poorly constrained biases \citep{lee:09}, and
quantities such as the star-formation rate (SFR) can differ greatly
depending on which observational tracer is used to estimate
them. Therefore, in order to interpret the zoo of galaxies selected at
different wavelengths (e.g. LBGs, EROs, DRGs, DOGs, BzKs, BM/BX, SMGs,
etc.), it is important to develop models that can accurately predict
observable quantities over the full range of wavelengths probed by
modern panchromatic surveys.

Important theoretical advances have been made in the past few years,
as well, with the development of radiative transfer (RT) codes that,
coupled with a model of the distribution of stars, gas, and dust in a
galaxy, can produce detailed panchromatic predictions of the galaxy's
appearance and photometric properties
\citep{silva:98,jonsson:06,jonsson06a,jonsson:10}. One approach is to use
idealized galaxy models (e.g. spheroid plus disc), coupled with a
radiative transfer code, as in \citet{silva:98}. Another is to
use hydrodynamic simulations to provide more detailed spatial and
morphological information, as in \citet{jonsson06a}.

While powerful, these tools are still computationally quite
expensive. Producing panchromatic predictions for statistical samples
of galaxies in a cosmological context remains beyond reach, without
some shortcuts. Semi-analytic models (SAMs) of galaxy formation, which
apply simple but physically motivated recipes for the physical
processes that shape galaxy formation, within the framework of
structure formation predicted by \LCDM\ ($\Lambda$ Cold Dark Matter),
can provide predictions of bulk galaxy properties (such as star
formation and chemical enrichment history, radial size, total stellar
mass or luminosity, ratio of spheroid to disc, etc.) for very large
numbers of galaxies. SAMs have been shown to reproduce many observed
properties of galaxies
\citep[e.g.][]{kwg:93,cafnz:94,sp:99,kauffmann:99,cole:00,spf:01}, and
to agree reasonably well with the results of numerical hydrodynamic
simulations in their predictions for basic quantities such as the rate
of accretion of cold gas and galaxy mergers
\citep{yoshida:02,cattaneo:07}. In particular, recent models that
include ``radio mode'' feedback from Active Galactic Nuclei (AGN)
reproduce quite well the global properties of massive galaxies over a
broad range of cosmic history
\citep[e.g.][]{croton:06,bower:06,menci:06,kang:06,monaco:07,s08},
although reproducing the properties of low-mass galaxies remains a
challenge \citep{fontanot:09,guo:10}.

Using a set of recipes for gas accretion and cooling, merging, star
formation, stellar feedback, chemical enrichment, and optionally black
hole growth and AGN feedback, a SAM outputs a distribution of ages and
metallicities for all the stars within the spheroid and disc
components of a galaxy (more details are given in
Section~\ref{sec:models}). This information is convolved with ``simple
stellar population'' (SSP) models \citep[e.g.][]{bruzual&charlot03},
which specify, for a given stellar Initial Mass Function (IMF), the
luminosity as a function of wavelength for a stellar population of a
given age and metallicity, in order to predict the unattenuated
Spectral Energy Distribution (SED) of starlight in the galaxy. The
predictions of stellar population models from different groups have
largely converged in their predictions, particularly in the UV and
optical, making this component of the modelling relatively
robust\footnote{The convergence is not as good in the NIR, where there
  are still significant uncertainties regarding the importance of
  contributions from Thermally Pulsating Asymptotic Giant Branch
  (TP-AGB) stars \citep{maraston:05}.}.

For the more difficult step of predicting how this starlight is
absorbed and re-radiated by dust, one possible approach is to couple
the predictions of a SAM directly with a radiative transfer code
\citep[e.g.][]{granato:00,baugh:05,fontanot:07}. Because SAMs are not
able to track the detailed internal structure or morphology of
galaxies, this requires the assumption of an idealized geometry such
as a spheroid plus disc (where the sizes and masses of the components
are specified by the SAM). However, even this approach is
prohibitively expensive for large numbers of galaxies, and also has
the disadvantage that the simplified geometries may not be
representative of the diversity of galaxy types, particularly for LIRG
or ULIRG-like objects, many of which are known to be merging
systems. Moreover, the dust models contain a large number of free
parameters, which must be tuned by fitting a chosen set of
observations, and may or may not be constant from galaxy to galaxy or
over cosmic time.

An alternative approach is to develop an analytic or semi-analytic
model to estimate the fraction of starlight that is absorbed by dust
in a given galaxy, based on its geometry, metal content and stellar
populations. Early SAMs
\citep{guiderdoni:87,lacey:93,guiderdoni:98,kauffmann:99,sp:99,devriendt:00}
approached this by assuming that the face-on B or V-band optical depth
of the dust in the disc is proportional to the column density of
metals in the gas phase, that the inclination dependence is that
predicted by a simple ``slab'' model in which the dust and stars are
uniformly mixed, and that the wavelength dependence is given by a
fixed ``attenuation law'', such as a Galactic extinction law or the
starburst attenuation law of
\citet{calzetti:00}. \citet{charlot&fall00} proposed a two-component
model that separately accounts for the extinction due to diffuse
``cirrus'' in the disc and that due to the dense ``birth clouds''
surrounding newly born stars. \citet{delucia:07} combined the two
approaches, using a ``slab'' model to treat the cirrus component and
adopting the \citet{charlot&fall00} model to treat the young stars
($\lesssim 10^{7}$ yr). \citet{fontanot:09a} tested a wide range of
such simple analytic approaches from the literature against full
radiative transfer, applied within the MORGANA \citep{monaco:07}
SAM. They concluded that bulk properties (such as UV, optical,
  and NIR luminosity functions) predicted by the SAMs using analytic
dust recipes agreed quite well with the results of the full radiative
transfer, at a fraction of the computational cost.

With an estimate for the total energy absorbed by dust in hand, and
assuming that all of this energy is re-radiated in the IR, one can use
observationally derived or observationally calibrated template SEDs
describing the wavelength dependence of the dust emission
\citep{devriendt:99,chary-elbaz:01,dale:02,lagache:04,rieke09}, or
modified Planck functions \citep{kaviani:03} to compute IR
luminosities
\citep{guiderdoni:98,devriendt:00,hatton:03,blaizot:04}. Observationally,
it is known that the mid- to far-IR colours (i.e. the ratio of warm to
cool dust) are correlated with the total IR luminosity
of the galaxy \citep{sanders:96}. Accordingly, models based on this
approach use an SED library indexed by the total IR luminosity; i.e.,
the total IR luminosity of the model galaxy is used to select the
appropriate FIR template. \citet{fontanot:10} compared this kind of
approach, again applied to the MORGANA SAM, with the results of
coupling the same SAM with the full RT model GRASIL
\citep{silva:98}. Again, the agreement for statistical quantities such
as luminosity functions was quite good.

Our goal in this paper is to develop fully semi-analytic models of
galaxy formation that can predict photometric properties of galaxies
from the far-UV to the far-IR with reasonable accuracy. To do this, we
adopt a modified version of the \citet{charlot&fall00} model to
estimate how much light is absorbed by dust in each galaxy, assume
that all of this energy is re-radiated by dust, and employ
observationally-calibrated templates to estimate luminosities in the
mid- to far-IR. We first ask how successfully this simple approach can
reproduce a compilation of observations spanning the FIR to the FUV
and a redshift range of zero to five. We expect such a na\"ive
approach to fail in some respects, and we attempt to identify the
physical lessons that we can draw from the points of failure. To aid
in this, we vary some of the more uncertain ingredients of our models
to study the effect on the observables. The resulting fiducial models
will be used to create mock catalogs for pan-chromatic surveys such as
CANDELS \citep{grogin:11,koekemoer:11}, and to make predictions of
  the Extragalactic Background Light (EBL) and its build-up over
  cosmic time. The implications for gamma ray observations will be
  explored in a companion paper (Gilmore, Somerville, Primack \&
  Dominguez 2011; hereafter GSPD).

This paper is structured as follows. In \S\ref{sec:models} we describe
the ingredients of the semi-analytic model, including our treatment of
dust attenuation and emission. In \S\ref{sec:results} we present
predictions for luminosity functions, counts, and related quantities
from the FUV to the sub-mm for several model variants, and compare
them with an extensive compilation of observations. We discuss and
summarize our results in \S\ref{sec:discussion}. 

%=======================
% 2
\section{Models}
\label{sec:models}
%=======================

\subsection{Semi-analytic Models}

\subsubsection{Galaxy Formation}
\label{sec:models:sam}

The semi-analytic models used here have been described in detail in
\citet{sp:99}, \citet{spf:01} and especially \citet[hereafter
  S08]{s08}, and we refer the reader to those papers for details. Here
we provide a brief summary of the basic ingredients of the SAM, which
include the growth of structure in the dark matter component in a
hierarchical clustering framework, radiative cooling of gas, star
formation, supernova feedback, AGN feedback, galaxy merging within
dark matter haloes, metal enrichment of the interstellar medium (ISM)
and intracluster medium (ICM), and the evolution of stellar
populations.

We assume a standard $\Lambda$CDM universe and a Chabrier IMF
\citep{chabrier:03}.  We consider two sets of cosmological parameters
here. One is the ``concordance'' cosmology (C-\LCDM), $\Omega_m$ =
0.3, $\Omega_{\Lambda}$ = 0.7, $H_0$ = 70.0, and $\sigma_8$ = 0.9,
which was used by S08 and has also been used by a large number of
other studies in the literature. We also present results for an
updated set of cosmological parameters that are consistent with the
five year WMAP results (WMAP5): $\Omega_m$ = 0.28, $\Omega_{\Lambda}$
= 0.72, $H_0$ = 70.0, $\sigma_8$ = 0.81, and $n_s=0.96$
\citep{komatsu:09}.  We note that these values are generally
consistent with those obtained from the analysis of the seven-year
WMAP data release \citep{komatsu10}. The adopted baryon fraction is
0.1658. We consider WMAP5 to be our ``fiducial'' model and present
results from C-\LCDM\ for ease of comparison with previous work.

The merging histories (or merger trees) of dark matter haloes are
constructed based on the Extended Press-Schechter formalism using the
method described in \citet{sk:99}, with improvements described in
S08. These merger trees record the growth of dark matter haloes via
merging and accretion, with each ``branch'' representing a merger of
two or more haloes. We construct grids of ``root halos'' spanning
  the range $V_{\rm vir}= 30-1200$ km/s, where $V_{\rm vir}$ is the
  circular velocity at the virial radius, and resolve the merger
  history of each root halo down to progenitors at least 0.01 times
  the mass of the root. For root halos more massive than $10^{12}\,
  M_{\odot}$, we follow the merger histories down to a minimum
  progenitor mass of $10^{10}\,M_{\odot}$. We have checked that our
  results are robust to the chosen mass resolution of the trees. 

Whenever dark matter haloes merge, the central galaxy of the largest
progenitor becomes the new central galaxy, and all others become
`satellites'. Satellite galaxies lose angular momentum due to
dynamical friction as they orbit and may eventually merge with the
central galaxy. To estimate this merger timescale we use a variant of
the Chandrasekhar formula from \citet{boylan-kolchin:08}. Tidal
stripping and destruction of satellites are also included as described
in S08. We have checked that the resulting mass function and radial
distribution of satellites (sub-haloes) agrees with the results of
high-resolution N-body simulations that explicitly follow
sub-structure \citep{maccio:10}.

Before the Universe is reionised, each halo contains a mass of hot gas
equal to the universal baryon fraction times the virial mass of the
halo. After reionisation, the photo-ionising background suppresses
the collapse of gas into low-mass haloes.  We use the results of
\citet{gnedin:00} and \citet{kravtsov:04} to model the fraction of
baryons that can collapse into haloes of a given mass after
reionisation, assuming that the universe was fully reionized by
$z=11$.

When a dark matter halo collapses, or experiences a merger that at
least doubles the mass of the largest progenitor, the hot gas is
shock-heated to the virial temperature of the new halo. This radiating
gas then gradually cools and collapses. The cooling rate is estimated
using a simple spherically symmetric model, based on the following
picture. Assuming that the density profile of the gas decreases
monotonically with increasing radius, and the cooling rate is more
rapid for dense gas, at any moment we can define the ``cooling
radius'' as the radius within which all the gas will have had time to
cool within a time $t_{\rm cool}$. Then, assuming that the initial
density profile of the gas is a singular isothermal sphere ($\rho_{\rm
  gas} \propto r^{-2}$), the cooling rate is given by:
\begin{equation}
\dot{m}_{\mathrm{cool}}=\frac{1}{2}m_{\mathrm{hot}}\frac{r_{\mathrm{cool}}}
{r_{\mathrm{vir}}}\frac{1}{t_{\mathrm{cool}}},
\end{equation}
where $m_{\mathrm{hot}}$ is the mass of the hot halo gas,
$r_{\mathrm{vir}}$ is the virial radius of the dark matter halo, and
$r_{\mathrm{cool}}$ is the cooling radius. We calculate the cooling
radius using the metallicity dependent atomic cooling curves of
\citet{sutherland:93}. Previous studies have used different values for
the cooling time $t_{\rm cool}$ (e.g., the Hubble time, the time since
the last halo merger, or the halo dynamical time). In the models of
S08, and also here, it is assumed to be equal to the halo dynamical
time, $t_{\rm dyn} \propto r_{\rm vir}/V_{\rm vir}$, where $V_{\rm
  vir}$ is the virial velocity of the halo.

In some cases the cooling radius can be formally larger than the
virial radius. In this case, the cooling rate is limited by the infall
rate:
\begin{equation}
\dot{m}_{\mathrm{cool}}=\frac{1}{2}m_{\mathrm{hot}}\frac{1}{t_{\mathrm{cool}}}.
\end{equation}
The cooling radius limited regime ($r_{\rm cool} < r_{\rm vir}$) is
often associated with what have been termed ``hot flows'' in
hydrodynamic simulations, in which gas is shock heated in a diffuse
halo and then cools. The infall limited cooling regime ($r_{\rm cool}
> r_{\rm vir}$) is associated with ``cold flows'', in which gas
streams into the halo along dense filaments, without ever getting
heated \citep{birnboim_dekel:03,dekel_birnboim:06,keres:05}.

In the present SAM, we assume that the cold gas is accreted only by
the central galaxy of the halo, but in reality satellite galaxies
should also receive some measure of new cold gas. In addition, we
assume that all newly cooling gas initially collapses to form a
rotationally supported disc. The scale radius of the disc is computed
based on the initial angular momentum of the gas and the halo profile,
assuming that angular momentum is conserved and that the self-gravity
of the collapsing baryons causes contraction of the matter in the
inner part of the halo \citep{blumenthal:86,flores:93,mo:98}. This
approach has been shown to reproduce the observed size versus stellar
mass relation for discs from $z\sim 0$--2 \citep{somerville:08}.

Star formation occurs in two modes, a ``quiescent'' mode in isolated
discs, and a merger-driven ``starburst'' mode. Star formation in
isolated discs is modelled using the empirical Schmidt-Kennicutt
relation \citep{kennicutt:98}, assuming that only gas above a fixed
critical surface density is eligible to form stars. The efficiency and
timescale of the merger driven burst mode is a function of the merger mass
ratio and the gas fractions of the progenitors, and is based on the
results of hydrodynamic simulations \citep{robertson:06a,hopkins:09a}.

Some of the energy from supernovae and massive stars is assumed to be
deposited in the ISM, resulting in the driving of a large-scale
outflow of cold gas from the galaxy. The mass outflow rate is
proportional to the star formation rate and inversely proportional to
the galaxy circular velocity (escape velocity) to the power of
$\alpha_{\rm rh}$, where $\alpha_{\rm rh}\simeq 2$, as expected for
``energy driven'' winds. Some fraction of this ejected gas escapes
from the potential of the dark matter halo, while some is deposited in
the hot gas reservoir within the halo, where it becomes eligible to
cool again. The fraction of gas that is ejected from the disc
but retained in the halo versus ejected from the disc and halo is a
function of the halo circular velocity (see S08 for details), such
that low-mass haloes lose a larger fraction of their gas.

Each generation of stars also produces heavy elements, and chemical
enrichment is modelled in a simple manner using the instantaneous
recycling approximation. For each parcel of new stars ${\rm d}m_*$, we
also create a mass of metals ${\rm d}M_Z = y \, {\rm d}m_*$, which we
assume to be instantaneously mixed with the cold gas in the disc. The
yield $y$ is assumed to be constant, and is treated as a free
parameter. When gas is removed from the disc by supernova driven winds
as described above, a corresponding proportion of metals is also
removed and deposited either in the hot gas or outside the halo,
following the same proportions as the ejected gas.

Mergers are assumed to remove angular momentum from the disc stars and
to build up a spheriod. The efficiency of disc destruction and
spheroid growth is a function of progenitor gas fraction and merger
mass ratio, and is parameterized based on hydrodynamic simulations of
disc-disc mergers \citep{hopkins:09a}. These simulations indicate that
more ``major'' (closer to equal mass ratio) and more gas-poor mergers
are more efficient at removing angular momentum, destroying discs, and
building spheroids. Note that the treatment of spheroid formation in
mergers used here has been updated relative to S08 as described in
\citet{hopkins:09b}. The updated model produces good agreement with
the observed fraction of disc vs. spheroid dominated galaxies as a
function of stellar mass.

In addition, mergers drive gas into galactic nuclei, fueling black
hole growth. Every galaxy is born with a small ``seed'' black hole
(typically $\sim 100\, \msun$ in our standard models). Following a
merger, any pre-existing black holes are assumed to merge fairly
quickly, and the resulting hole grows at its Eddington rate until the
energy being deposited into the ISM in the central region of the
galaxy is sufficient to significantly offset and eventually halt
accretion via a pressure-driven outflow. This results in
self-regulated accretion that leaves behind black holes that naturally
obey the observed correlation between BH mass and spheroid mass or
velocity dispersion \citep{dimatteo:05,robertson:06b,s08}.

There is a second mode of black hole growth, termed ``radio mode'',
that is thought to be associated with powerful jets observed at radio
frequencies. In contrast to the merger-triggered mode of BH growth
described above (sometimes called ``bright mode'' or ``quasar mode''),
in which the BH accretion is fueled by cold gas in the nucleus, here,
hot halo gas is assumed to be accreted according to the Bondi-Hoyle
approximation \citep{bondi:52}. This leads to accretion rates that are
typically only about $\la 10^{-3}$ times the Eddington rate, so that
most of the BH's mass is acquired during episodes of ``bright mode''
accretion. However, the radio jets are assumed to couple very
efficiently with the hot halo gas, and to provide a heating term that
can partially or completely offset cooling during the ``hot flow''
mode (we assume that the jets cannot couple efficiently to the cold,
dense gas in the infall-limited or cold flow regime). This ``radio
mode feedback'' appears to be able to quite successfully solve many of
the problems experienced by previous generations of \LCDM-based galaxy
formation models; for example, the excess number density and overly
high specific star formation rates and blue colours of massive galaxies
\citep{croton:06,bower:06,s08}.

For each galaxy, we store a two-dimensional grid of the mass of stars
with a given age and metallicity, separately for the disc and
spheroid. We then convolve this distribution with the predictions of
the simple stellar population (SSP) models of \citet{bruzual&charlot03} to
obtain the SED of the unattenuated starlight. We use the models based
on the Padova 1994 isochrones with a \citet{chabrier:03} IMF.

\subsubsection{Dust Attenuation}
\label{sec:models:dustatt}

Our model for dust extinction is based on the model proposed by
\citet{charlot&fall00}.  As in their model, we consider extinction by
two components, one due to the diffuse dust in the disc and another
associated with the dense `birth clouds' surrounding young star
forming regions. The $V$-band, face-on extinction optical depth of the
diffuse dust is given by
\begin{equation}
\tau_{V,0} = \tau_{\mathrm{dust,0}}\, Z_{\mathrm{cold}}\, m_{\mathrm{cold}}/
(r_{\mathrm{gas}})^2,
\end{equation}
where $\tau_{\mathrm{dust,0}}$ is a free parameter,
$Z_{\mathrm{cold}}$ is the metallicity of the cold gas,
$m_{\mathrm{cold}}$ is the mass of the cold gas in the disc, and
$r_{\mathrm{gas}}$ is the radius of the cold gas disc, which is
assumed to be a fixed multiple of the stellar scale length (see
S08). To compute the actual extinction we assign a random inclination
to each disc galaxy and use a standard `slab' model; i.e. the extinction in
the $V$-band for a galaxy with inclination $i$ is given by:
\begin{equation}
A_V = -2.5 \log_{10}\left[\frac{1-\exp[-\tau_{V,0}/\cos(i)]}{\tau_{V,0}/\cos(i)}\right].
\end{equation} 

Additionally, stars younger than $t_{\rm BC}$ are enshrouded in a
cloud of dust with optical depth
$\tau_{\mathrm{BC,V}}=\mu_{\mathrm{BC}}\, \tau_{V,0}$, where we treat
$t_{\rm BC}$ and $\mu_{\mathrm{BC}}$ as free parameters.  Finally, to
extend the extinction estimate to other wavebands, we assume a
starburst attenuation curve \citep{calzetti:00} for the diffuse dust
component and a power-law extinction curve
$A_{\lambda}\propto(\lambda/5500\AA)^n$, with $n=0.7$, for the birth
clouds \citep{charlot&fall00}. 

\subsubsection{Dust Emission}
\label{sec:models:dustem}

\begin{figure*} 
\begin{center}
\includegraphics[width=6.5in]{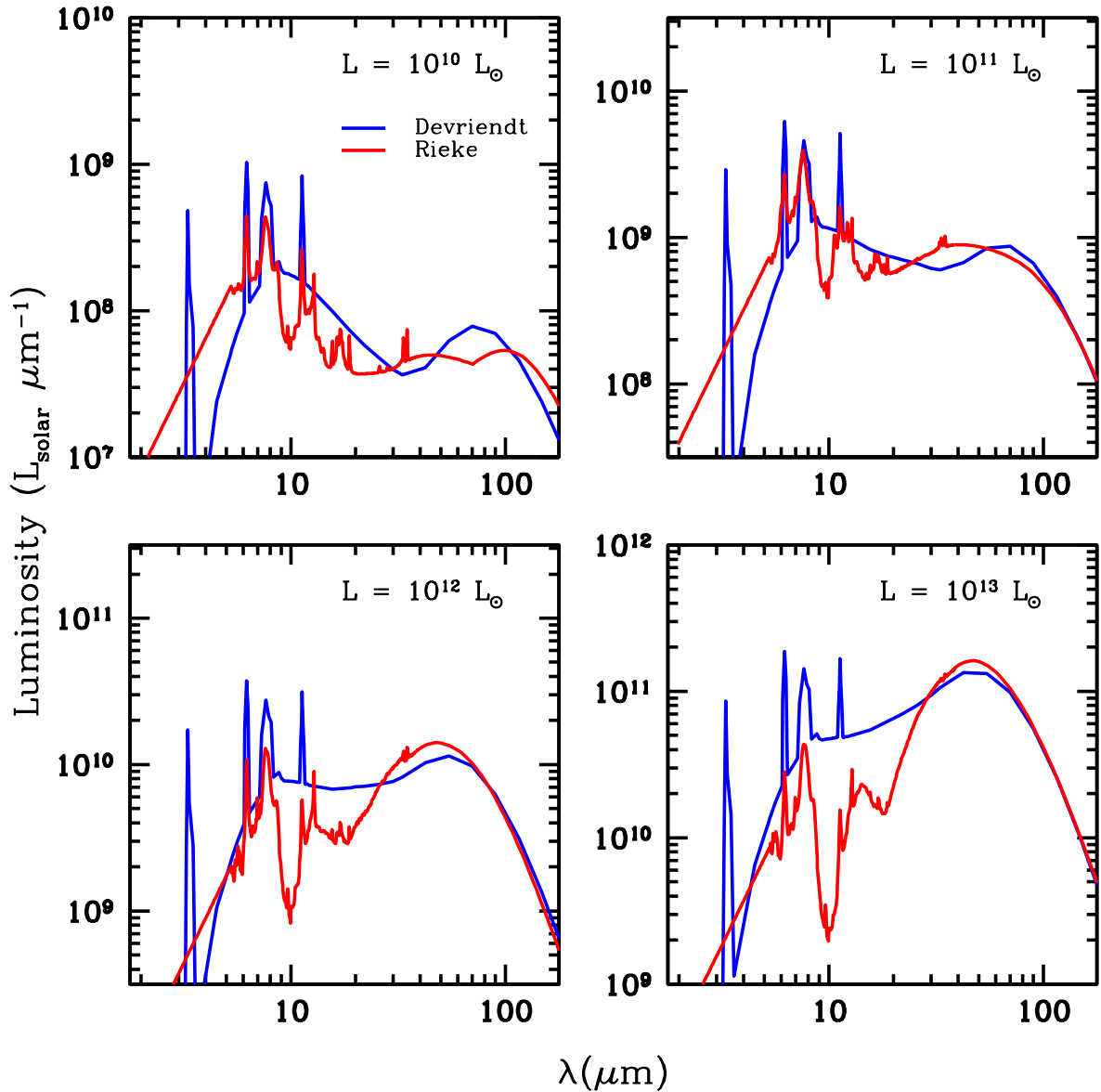}
\end{center}
\caption{Comparison of the dust emission templates of \citet{rieke09}
  (red) and \citet{devriendt:99} (blue).  The four panels show the
  dust emission templates used in this work for bolometric IR
  luminosities of 10$^{10}\, L_\odot$, 10$^{11} \, L_\odot$ (a LIRG),
  10$^{12} \, L_\odot$ (a ULIRG), and 10$^{13} \, L_\odot$ (an extremely
  IR-bright `Hyper--LIRG').
\label{fig:irtemp}}
\end{figure*}

Using the approach described above, we can compute the total fraction
of the energy emitted by stars that is absorbed by dust, over all
wavelengths, for each galaxy. We then assume that all of this absorbed
energy is re-radiated in the infra-red (we neglect scattering), and
thereby compute the total IR luminosity of the galaxy $L_{\rm IR}$. We
make use of dust emission templates to determine the SED of the dust
emission, based on the hypothesis that the shape of the dust SED is
well-correlated with $L_{\rm IR}$. The underlying physical notion is
that the distribution of dust temperatures is set by the intensity of
the local radiation field; thus more luminous or actively star forming
galaxies should have a larger proportion of warm dust, as is in fact
observed \citep{sanders:96}.

There are two basic kinds of approaches for constructing these sorts
of templates. The first is to use a dust model along with either
numerical or analytic solutions to the standard RT equations to create
a library of templates, calibrated by comparison with local
prototypes. This approach was pioneered by \citet{desert:90}, and has
been followed by many other workers
\citep[e.g.][]{guiderdoni:98,devriendt:99,devriendt:00}. \citet{desert:90}
posited three main sources of dust emission: polycyclic aromatic
hydrocarbons (PAHs), very small grains and big grains.  The grains are
composed of graphite and silicates, with small and big grains probably
dominated by graphite and silicate respectively.  The thermal
properties of each species are determined by the size distribution and
thermal state. Big grains are assumed to be in near thermal
equilibrium, and their emission can be modeled as a modified
black-body spectrum. However, small grains and PAHs are probably in a
state that is intermediate between thermal equilibrium and single
photon heating. They are therefore subject to temperature fluctuations
and their emission spectra are much broader than a modified black-body
spectrum. The detailed size distributions are modeled using free
parameters, which are calibrated by requiring the model to fit a set
of observational constraints, such as the extinction/attenuation
curves, observed IR colours and the IR spectra of local galaxies. Here
we make use of the templates derived by \citet[][hereafter
  DGS99]{devriendt:99} using a similar approach to \citet{desert:90}.

The second approach is to make direct use of observed SEDs for a set
of prototype galaxies
\citep[e.g.][]{chary-elbaz:01,dale:02,lagache:04}. We also make use of
the empirical SED templates recently published by \citet[hereafter
  R09]{rieke09}. They constructed detailed SEDs from published ISO,
IRAS and NICMOS data as well as previously unpublished IRAC, MIPS and
IRS observations. They modeled the far infrared SEDs assuming a single
blackbody with wavelength-dependent emissivity.  
The R09 library includes fourteen SEDs covering the $5.6 \times 10^9
L_\odot<L_{IR}< 10^{13} L_\odot$ range. Examples of the DGS99 and R09
SED templates are shown in Figure~\ref{fig:irtemp}.

The R09 templates have less emission in the PAH and mid-IR regions
than those of DGS99, particularly at the brightest luminosities.  The
R09 templates are also considerably more detailed in their
representation of PAH emission.  Being observationally based, the
shortest wavelengths predicted by the R09 templates are contaminated
by emission from direct starlight.  We have attempted to remove this
component by subtracting from each template the average amount of
starlight in the SED for galaxies of that IR luminosity in the local
universe.  The R09 templates also end abruptly at 5 $\mu$m, and we
have smoothed the transition to the shorter wavelength starlight
regime by extrapolating using a power-law of slope $\sim
\lambda^{3}$. Our results are not very sensitive to the choice of this
power-law slope.

\subsubsection{Galaxy Formation Parameters}

\begin{figure*} 
\begin{center}
\plottwo{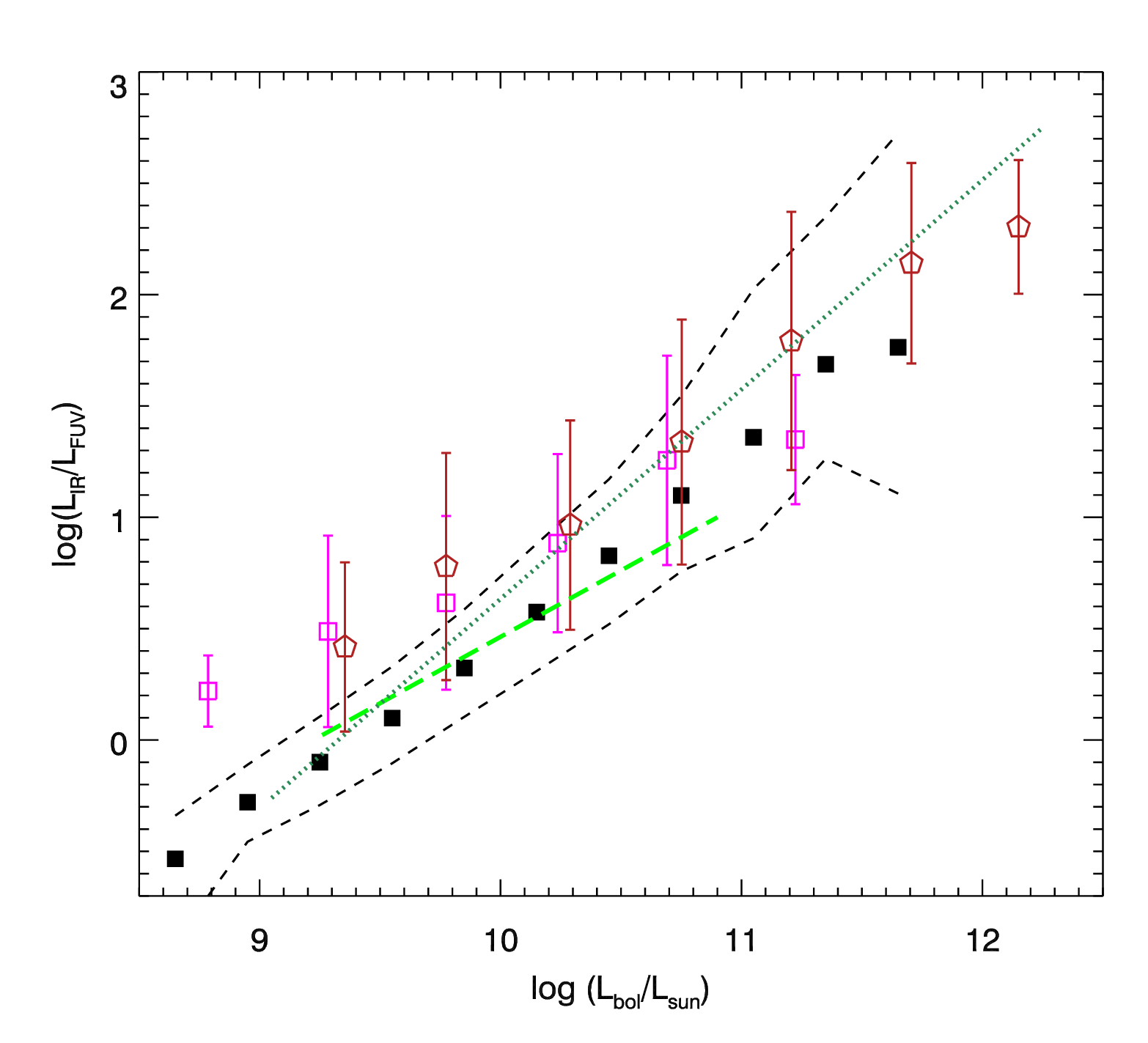}{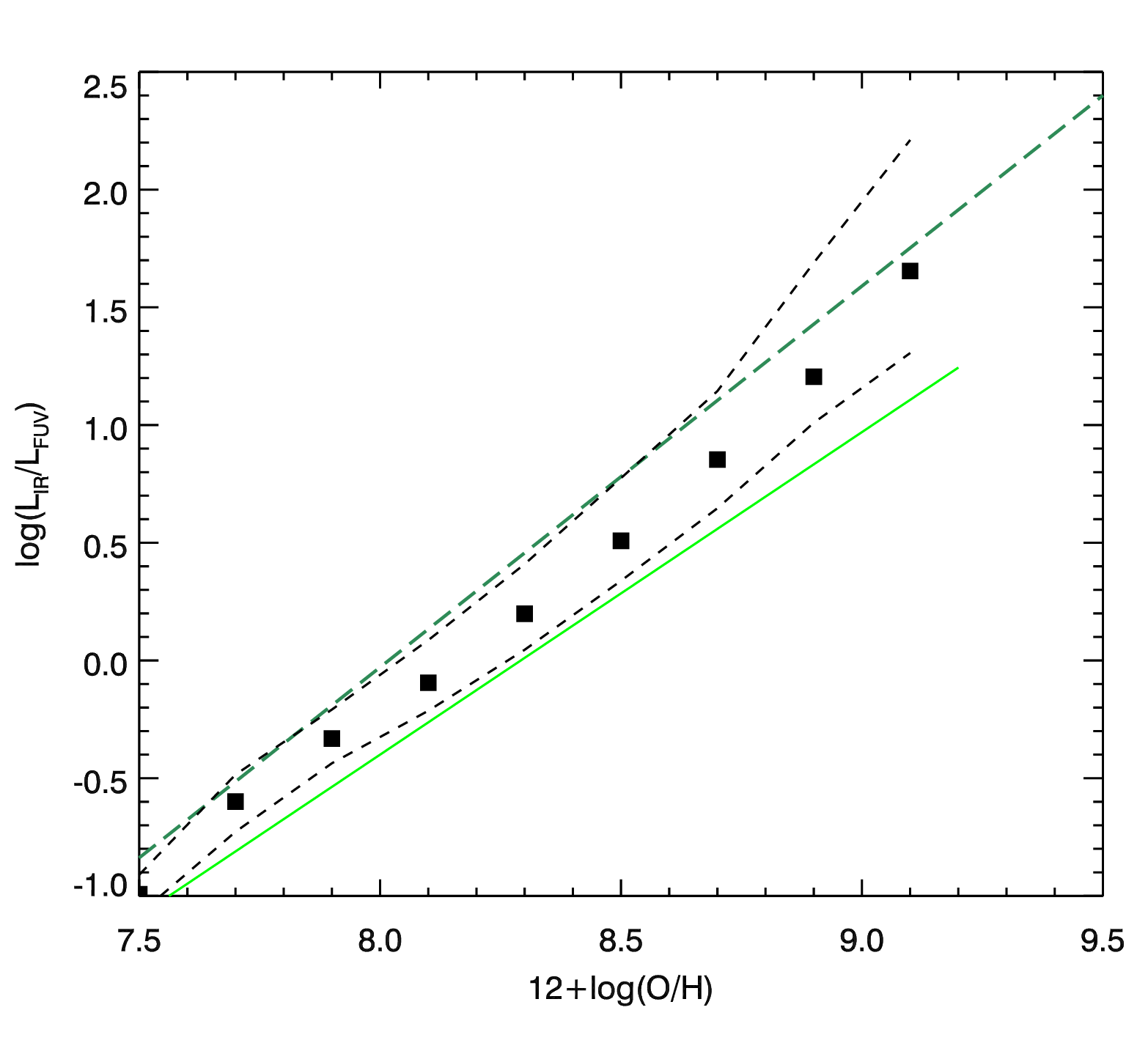}
\end{center}
\caption{\small Left: Galaxy luminosity in the IR relative to the
  luminosity in the UV vs. bolometric luminosity. Open symbols with
  error bars are observational estimates of this relationship for
  nearby galaxies from \protect\citet{buat:07}. The long dashed green
  line is the observational relation from \protect\citet{bell:03}, and
  the dotted line is the observational relation from
  \protect\citet{xu:06}. Right: Galaxy luminosity in the IR relative to
  the UV vs. metallicity of the cold gas in the galaxy. The dashed
  (green) line is the observational relation of
  \protect\citet{heckman:98} and the solid green line is the relation of
  \protect\citet{cortese:06}. In both panels, solid black squares show the
  medians for our fiducial model, and dashed black lines show the 16th
  and 84th percentiles.
\label{fig:fdust}}
\end{figure*}

The galaxy formation models contain a number of free parameters which
are tuned using observational constraints. A full list of the physical
parameters in the semi-analytic model is given in S08 (Table 2). The
most important parameters for the quantities presented in this paper
are those controlling supernova and AGN feedback; specifically, the
efficiency of supernova-driven outflows ($\epsilon_{\rm SN}^0$) and
its dependence on galaxy circular velocity ($\alpha_{\rm rh}$), and
the efficiency of heating by ``radio mode'' AGN feedback ($\kappa_{\rm
  radio}$). As in S08, the values of these parameters are adjusted in
order to obtain a good match to the observed $z\sim0$ stellar mass
function. The low-mass end of the stellar mass function is insensitive
to the AGN feedback recipe and is mainly controlled by the supernova
feedback recipe, while the reverse is true for the high-mass end. The
effective yield used for chemical evolution is fixed by matching the
zero-point of the galaxy stellar mass-metallicity relation (see
S08). The values of the parameters used in the C-\LCDM\ model
presented in this paper are identical to those for the C-\LCDM\ model
in S08, except that we have slightly increased the strength of
  the supernovae feedback ($\epsilon_{\rm SN}^0=1.5$ and $\alpha_{\rm rh}=2.5$), and
  decreased the efficiency of the radio mode feedback ($\kappa_{\rm
    radio}=2.5\times 10^{-3}$.)
For the fiducial WMAP5 model presented here, we have adjusted
the galaxy formation parameters slightly to account for the modified
cosmology, but they are quite similar to the parameters for the WMAP3
model presented in S08.

\subsubsection{Dust Parameters}

We also have three additional parameters that control the dust
attenuation in our model: the normalization of the face-on V-band
optical depth $\tau_{\mathrm{dust,0}}$, the opacity of the birthclouds
relative to the cirrus component $\mu_{\mathrm{BC}}$, and the time
that newly born stars spend enshrouded in their birthclouds, $t_{\rm
  BC}$. We first set $\tau_{\mathrm{dust,0}}$ by matching the
normalization of the observed relationship between $L_{\rm
  dust}/L_{\rm UV}$ vs. bolometric luminosity $L_{\rm bol}$, where
$L_{\rm dust}$ is the total luminosity absorbed by dust and re-emitted
in the mid- to far-IR and $L_{\rm UV}$ is the luminosity in the Far UV
($\sim1500 \AA$).  The predicted median relation and 1-$\sigma$
scatter is shown in Figure \ref{fig:fdust}, along with several
observational estimates. We also show the predicted relationship
between the cold gas metallicity and $L_{\rm dust}/L_{\rm UV}$
compared with observations in Figure \ref{fig:fdust}. We find good
agreement with $\tau_{\mathrm{dust,0}}=0.2$. With this value, we also
obtain good agreement with the observed optical through NIR luminosity
functions at $z=0$ (see Section~\ref{sec:results:lf}).

The birthcloud parameters $\mu_{\mathrm{BC}}$ and $t_{\rm BC}$ mainly
control the attenuation of UV light relative to longer wavelengths. At
$z=0$, the $g$ through $K$-band luminosity functions are insensitive
to the birthcloud parameters, while the FUV through u-bands are quite
dependent on them. We adjust these parameters in order to match the
$z=0$ FUV and NUV observed luminosity functions, finding good
agreement with $\mu_{\mathrm{BC}}=4.9$ and $t_{\rm BC}=2\times 10^{7}$
yr.

It remains an open question whether the properties of interstellar
dust have evolved over cosmic time and, if so, what impact this might
have on the appearance of high redshift galaxies. Recent work has
suggested that, at fixed $L_{\rm bol}$, galaxies may be less
extinguished at high redshift than one would expect if the relation
shown in Figure \ref{fig:fdust} were constant over all times
\citep[e.g.][]{reddy06,reddy:10}. In addition, we find, in agreement
with some previous studies \citep{lofaro:09,guo-white:09}, that if we
keep the dust parameters fixed at constant values, we underpredict the
number of UV-bright galaxies at high redshift. Therefore we
  introduce an ad hoc redshift dependence into the dust parameters,
  which we use in our fiducial model and most of its variants:
$\tau_{\mathrm{dust,0}}(z) = \tau_{\mathrm{dust,0}}(z=0)/(1+z)$, and
both $\mu_{\mathrm{BC}}$ and $t_{\rm BC}$ scale with $z^{-1}$ above
$z=1$. We chose this dependence because it allows us to fit the
  bright end UV and B-band luminosity functions at all redshifts where
  they are well constrained observationally. We also show results from
  a model with dust parameters that do not vary with redshift (``fixed
  dust'' model). 

\subsubsection{Model Variants}

\begin{table*}
\centering
\caption{Summary of Models}
\begin{tabular}{lccccr}
\hline \hline
name & cosmology  & dust attenuation & dust emission & dust parameters & designated line style\\
\hline \hline
WMAP5 fiducial & WMAP5 & composite & Rieke et al. (2009) & evolving & solid black \\
WMAP5+fixed dust & WMAP5 & composite & Rieke et al. (2009) & fixed & dash-dotted purple \\
WMAP5+Calzetti & WMAP5 & Calzetti & Rieke et al. (2009) & fixed & dashed red \\
WMAP5+DGS & WMAP5 & composite & Devriendt et al. (1999) & evolving & long-dashed blue \\
C-\LCDM & C-\LCDM & composite & Rieke et al. (2009) & evolving & dotted black\\
\hline \hline
\end{tabular}
\label{tab:models}
\end{table*}

We vary several of the uncertain ingredients of our models in order to
study the sensitivity of our results to these assumptions. As we have
already discussed, we consider two cosmologies, the ``concordance''
(C-\LCDM) or WMAP1 cosmology and the currently favored WMAP5
cosmology. We consider a model variant in which instead of using the
composite dust attenuation model (with cirrus plus birth clouds) based
on \citet{charlot&fall00}, we instead use a fixed attenuation curve
from \citet{calzetti:00}. We consider two different sets of empirical
dust emission SED templates (R09 and DGS99). As discussed above, we
also consider a model in which the dust parameters do not evolve
with redshift. This results in five models, which are summarized in
Table~\ref{tab:models}.  This table also specifies the line style that
will be used for each model in the plots of the following
section. For some quantities, some of the models produce the same
  predictions, in which case we show a subset of the models.

%=======================
% 3
\section{Results}
\label{sec:results}
%=======================

\begin{figure*} 
\begin{center}
\plottwo{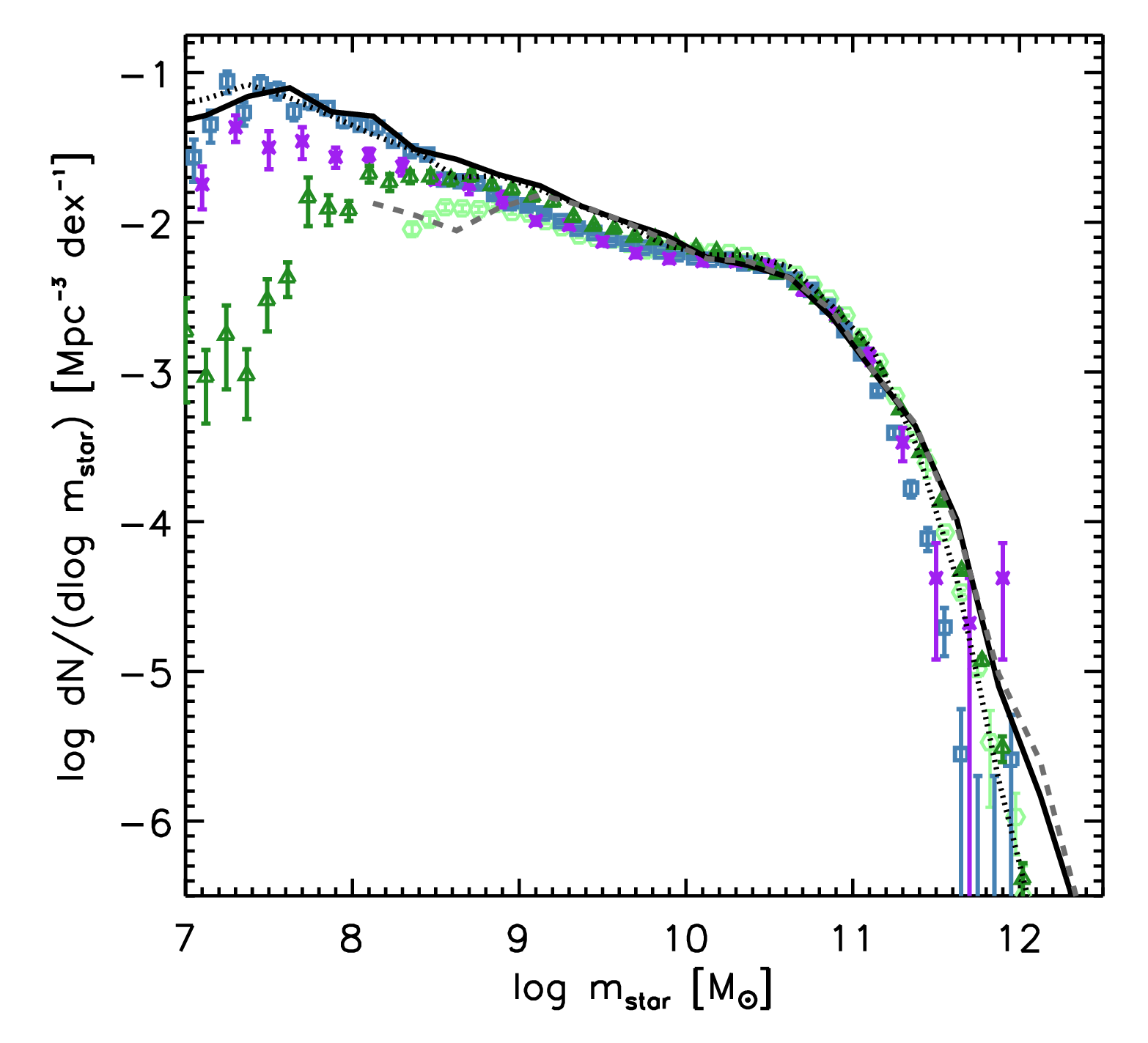}{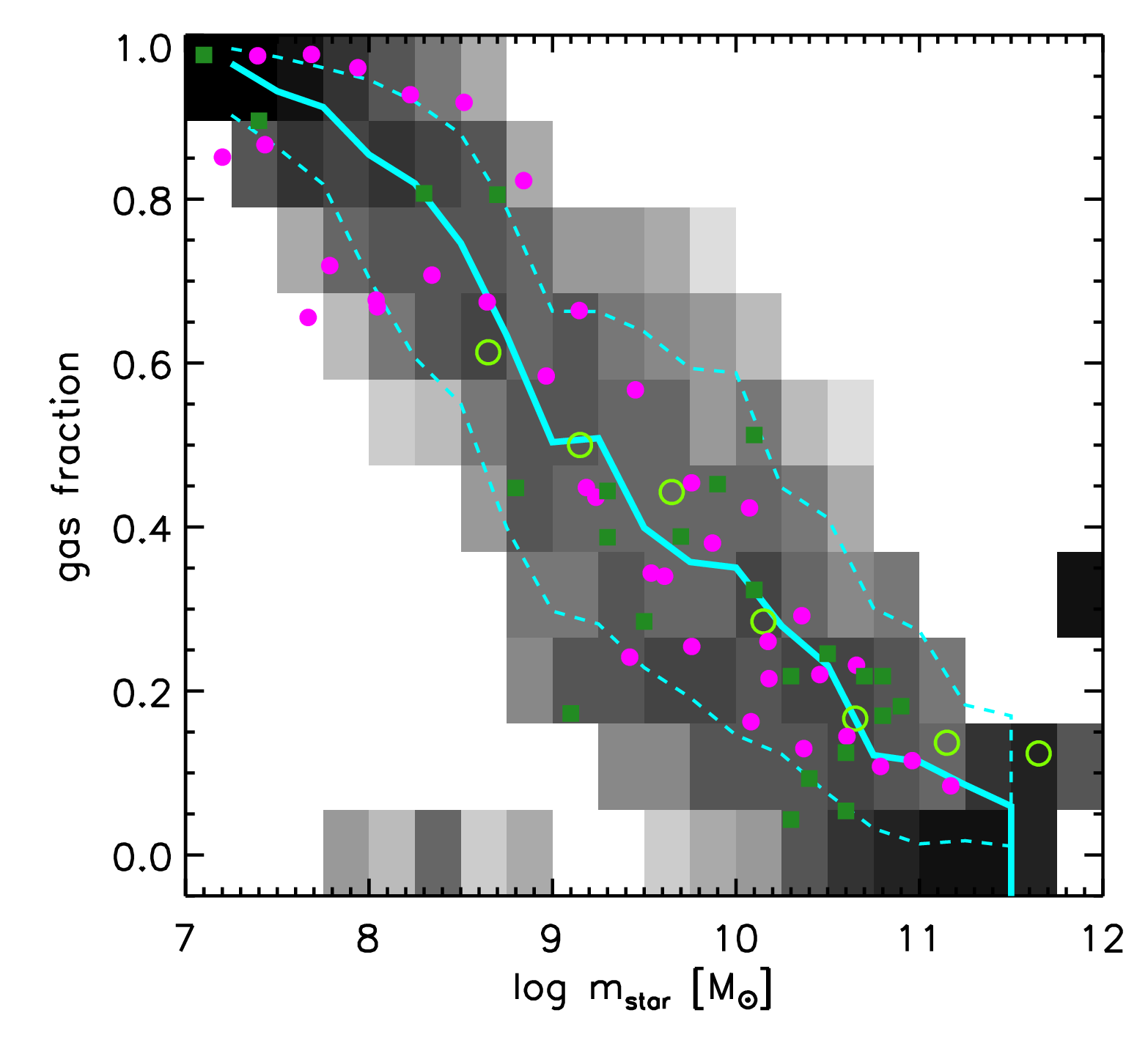}
\end{center}
\caption{\small Left: Stellar mass function at $z=0$. The solid line
  shows the fiducial (WMAP5) model, dotted line shows the
  C-\LCDM\ model, and the dashed gray line shows the fiducial model at
  the resolution of the simulations presented in S08. Symbols show the
  observationally derived stellar mass functions from
  \protect\citet[][blue squares]{baldry:08}, \protect\citet[][purple
    crosses]{baldry:11}, \protect\citet[][dark green
    triangles]{panter:07}, and \protect\citet[][light green
    hexagons]{li:09}. Note that the quoted values for the observed
  mass functions at $m_{\rm star} \lesssim 10^{8.5} M_{\odot}$ are
  likely to be lower limits, due to surface brightness selection
  effects. Right: Gas fraction for central disk-dominated galaxies in
  the fiducial model (gray shading as in S08). Solid and dashed lines
  show model median and 16 and 84th percentiles. Large open circles
  show the observations of \protect\citet{kannappan:04}. Filled
  squares show gas fractions from galaxies in the THINGS survey
  \protect\citep{leroy:08}, and small filled circles show observations
  from \protect\citet{baldry:08}.
\label{fig:mfstar}} 
\end{figure*}

\begin{figure*} 
\begin{center}
\plottwo{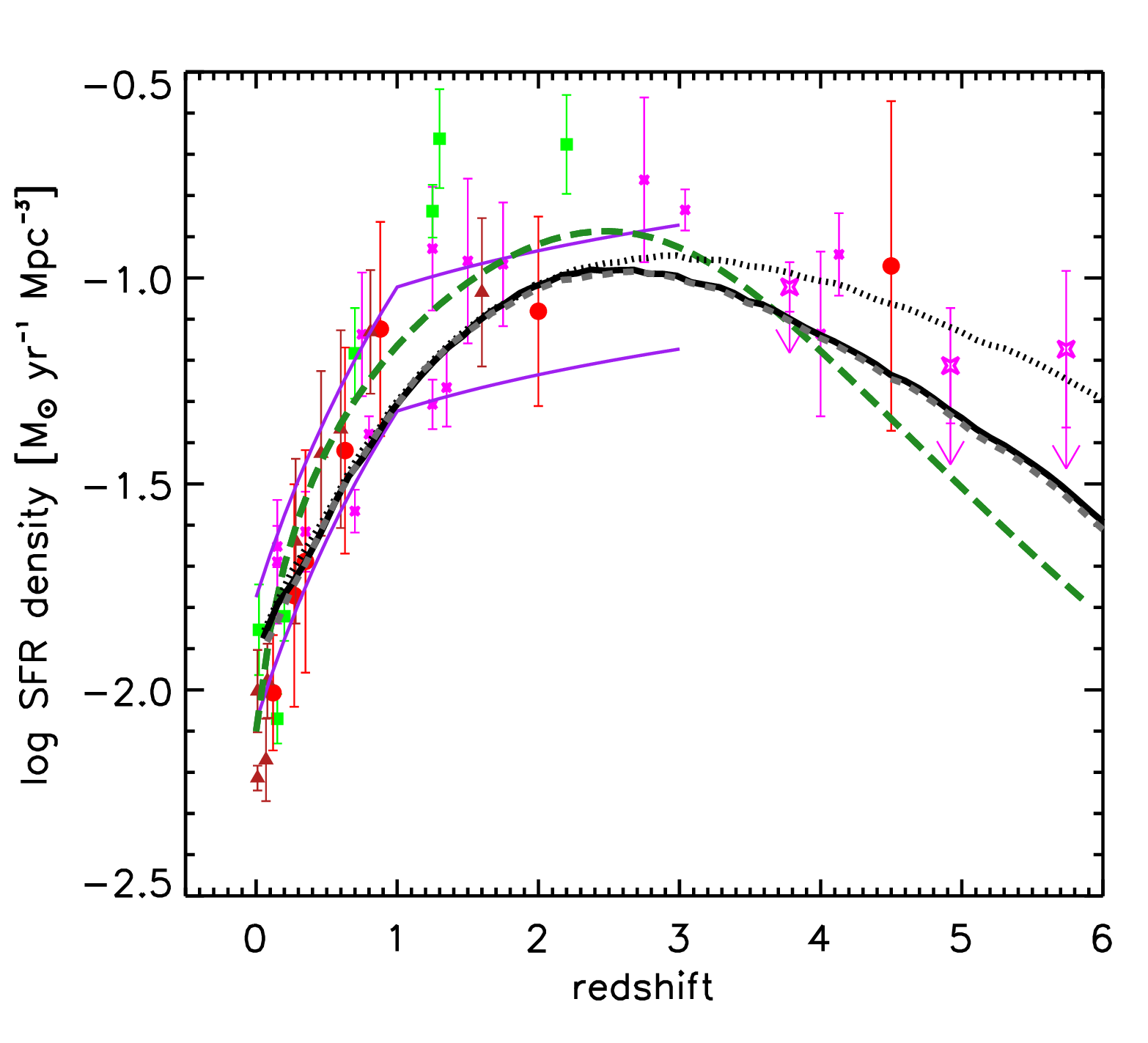}{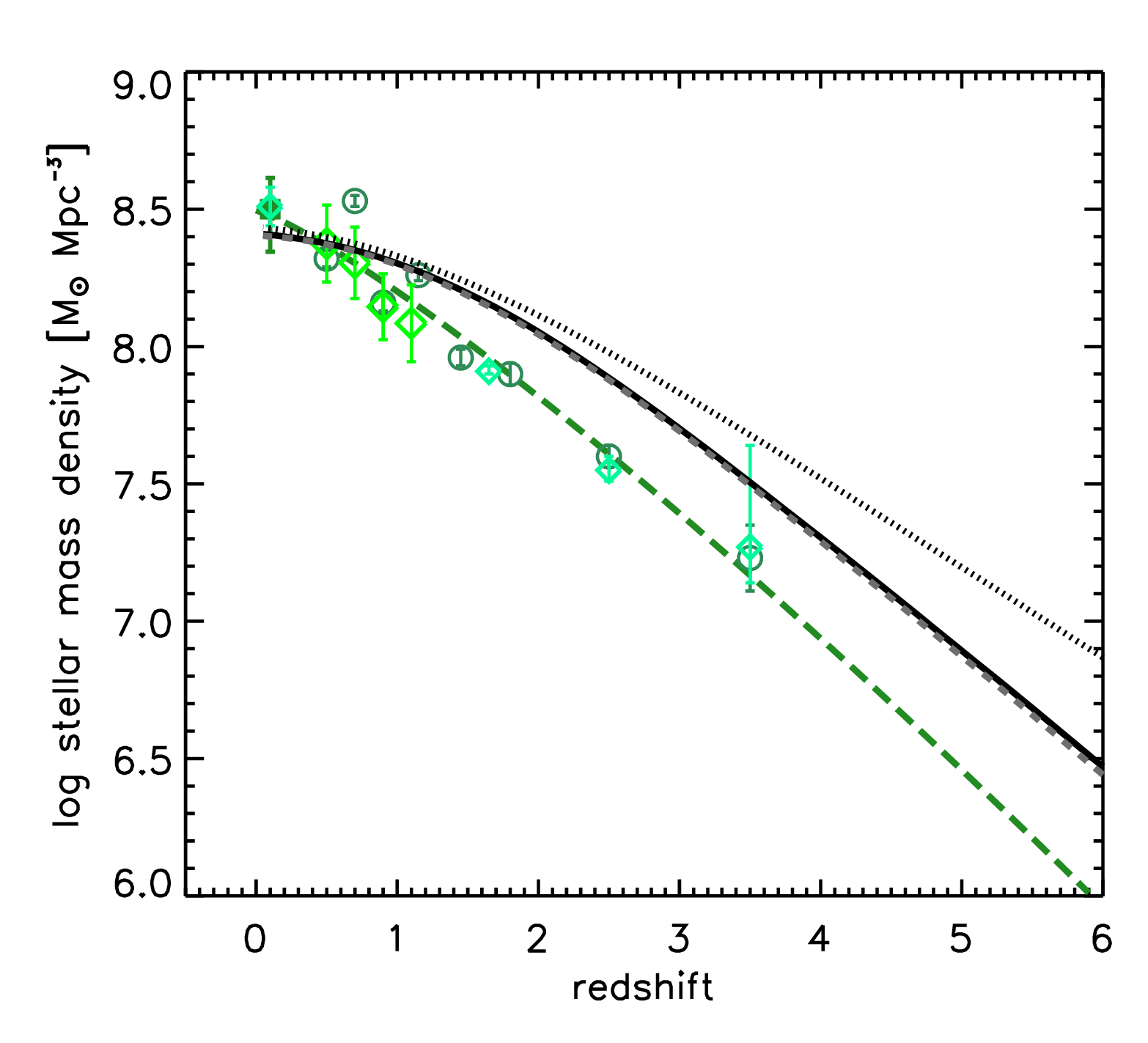}
\end{center}
\caption{\small Left: Global star formation history. Right: global
  stellar mass assembly history. Both panels: solid black lines show
  the predictions of the WMAP5 model; dotted lines show the
  C-\LCDM\ model; dark gray dashed lines show the fiducial model at
  the resolution of the simulations presented by S08.  Symbols show a
  compilation of observational estimates (references given in
  S08). The long-dashed lines are the observational estimates from
  \protect\citet{hopkins_beacom:06}.
\label{fig:sfhist}} 
\end{figure*}

\subsection{Luminosity Functions and Luminosity Densities}
\label{sec:results:lf}
\begin{figure*} 
\begin{center}
\includegraphics[width=6.5in]{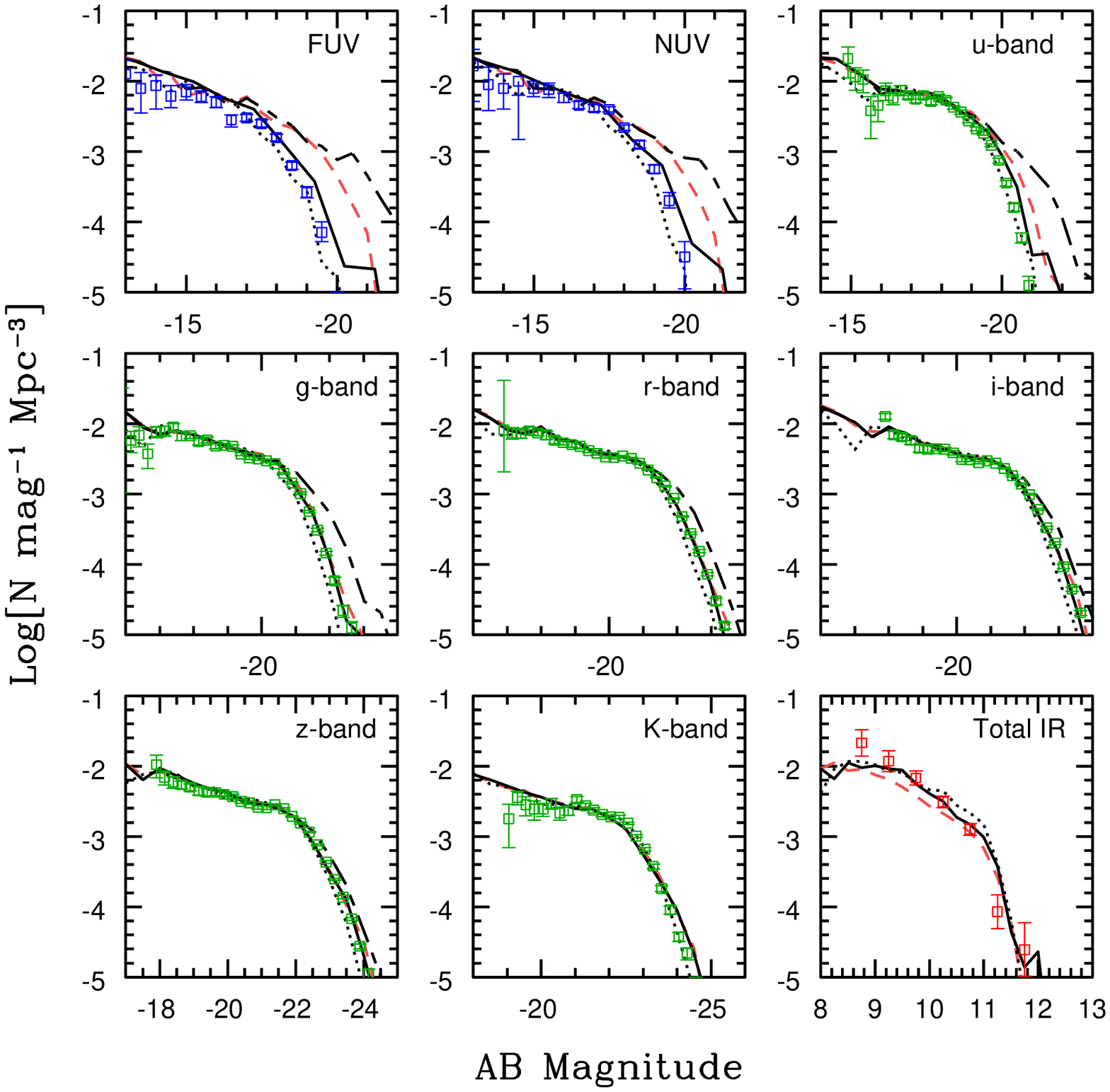}
\end{center}
\caption{Luminosity functions at $z=0$ in the FUV, NUV, SDSS ugriz and
  K bands, as well as the total IR (8-1000 $\mu$m).  The solid black
  line is our fiducial WMAP5 model using the composite dust recipe
  (Charlot-Fall), and dashed red shows the model with a single
  component dust model (Calzetti).  The dotted black line shows the
  C-\LCDM\ model. The black long-dashed line shows the
  predictions of the fiducial model without dust attenuation.
Data are from \citet{wyder05} (blue points,
  $\langle$z$\rangle$=0.05), \citet{bell03} (green points), and
  \citet{rodighiero10} (red points, $\langle$z$\rangle$=0.15).  For
  the total IR panel, the x-axis shows the logarithmic luminosity in
  solar units, and the y-axis has units of N dex$^{-1}$ Mpc$^{-3}$;
  all other axes are as indicated. 
\label{fig:lfz0}}
\end{figure*}

Although our models are very similar to those presented in S08,
  the resolution used here is much higher than the resolution used in
  the simulations presented in S08. In S08, halos were followed down
  to a minimum mass of $10^{10} M_{\odot}$, while here, root halos at
  $z=0$ are followed down to a minimum progenitor mass of $1.3 \times
  10^8 M_{\odot}$. Therefore we first investigate whether changing the
  resolution of the merger trees has any effect on our results. In
  Fig.~\ref{fig:mfstar} we show the two main quantities that we use to
  normalize our models, the stellar mass function and the gas fraction
  of disk-dominated galaxies as a function of stellar mass, at
  $z=0$. In S08, we stated that galaxies with stellar masses above
  $\sim 10^9 M_{\odot}$ should be reliably resolved. We see from
  Fig.~\ref{fig:mfstar} that the results above this mass change
  negligibly when we increase the mass resolution by almost two orders
  of magnitude. In addition, with no tuning, our model fits the
  observed stellar mass function extremely well down to the smallest
  masses where it has been measured. The gas fractions also appear to
  match observations down to very low mass objects ($\sim 10^7
  M_{\odot}$). 

Next, as a further orientation to our models, we show in
Fig.~\ref{fig:sfhist} the global star formation rate density and
stellar mass density across cosmological time. The global star
formation histories are similar in the two models below redshift two,
but the C-\LCDM~model has more early star formation because of the
larger amount of small-scale power. We can see that there is a
  very small difference between the fiducial model at the resolution
  of S08 and with the higher resolution. This shows that the
  resolution adopted by S08 was sufficient to resolve the galaxies
  that contribute the bulk of the global star formation rate and
  stellar mass density from $z\sim6$ to the present.

As a first test we investigate the luminosity functions at $z=0$ from
the rest-UV to the IR. In Figure~\ref{fig:lfz0}, we show the FUV and
NUV luminosity functions from GALEX, ugriz LFs from SDSS, the K-band
LF from 2MASS, and the total IR luminosity function (references given
in figure caption). Our fiducial model agrees very well with all of
these data, other than slightly overpredicting the numbers of the
  faintest galaxies in the GALEX FUV and NUV bands.
We see that the model
with a fixed attenuation curve (Calzetti model) is unable to
simultaneously reproduce the UV through optical bands, while the
composite (modified Charlot \& Fall model) does this well. In this
model, the young stars that produce most of the UV light are heavily
enshrouded in dense birth clouds, resulting in an effectively
age-dependent extinction curve.

\begin{figure*} 
\begin{center}
\includegraphics[width=6.5in]{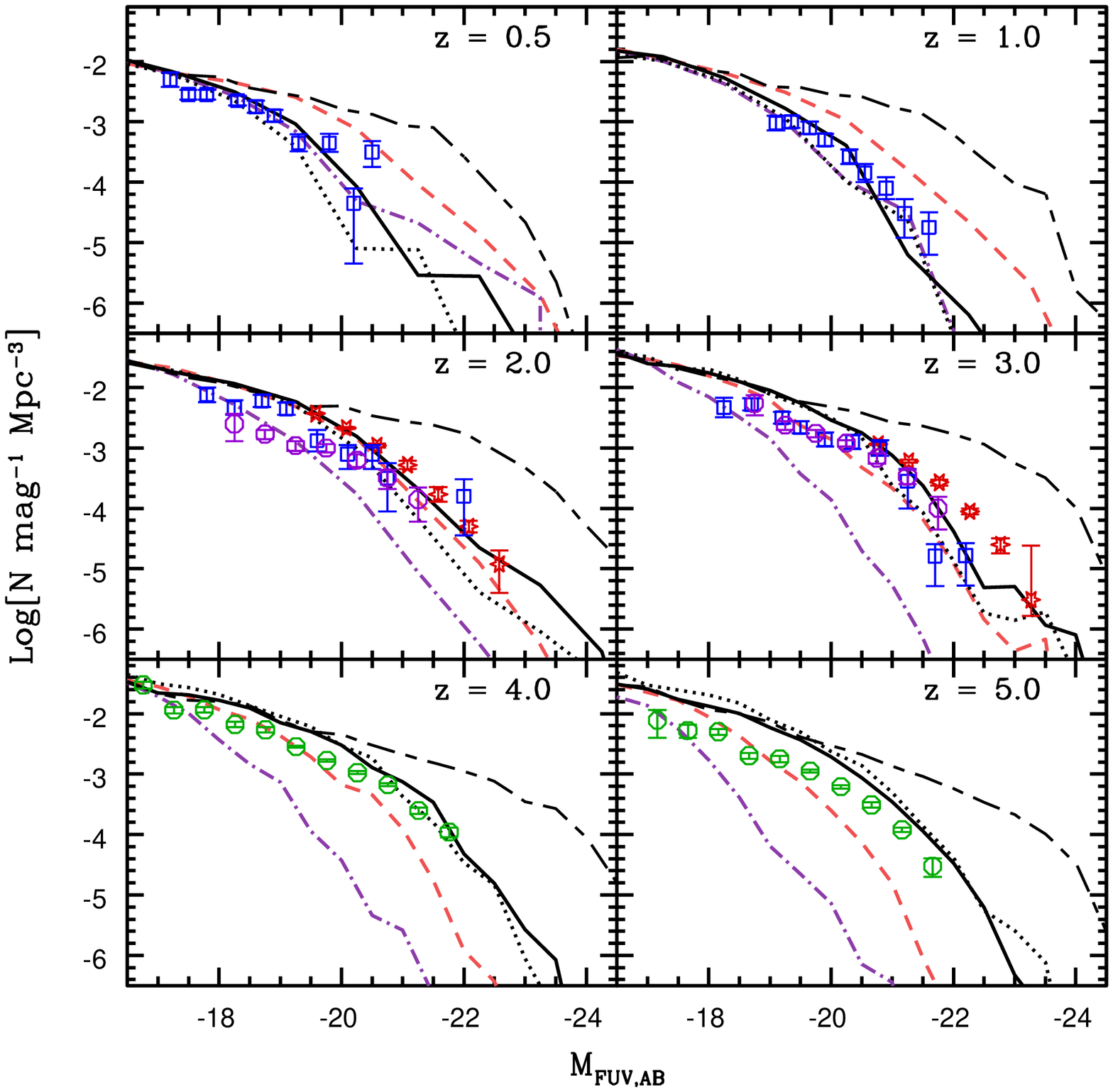}
\end{center}
\caption{\small Luminosity functions in the rest frame 1500 \AA\ UV
  band at several nonlocal redshifts.  The solid black line is our
  fiducial WMAP5 model using the evolving composite dust recipe
  (Charlot-Fall), the dash-dotted purple shows the composite dust model
  with fixed parameters, and dashed red is with the single component
  dust model (Calzetti).  The black long-dashed line shows the
  predictions of the fiducial model without dust attenuation. Blue
  squares are data from \citet{arnouts05}, violet circles are from
  \citet{hathi10}, red stars are from \citet{reddy08}, and green
  circles are from \citet{bouwens07}.
\label{fig:lfzuv}}
\end{figure*}

Next we explore the evolution over cosmic time of luminosity functions
in several wavebands. All luminosity functions are shown in the
rest-frame at the indicated redshift. We assume that the observations
have been properly completeness corrected and make no attempt to
determine whether our model galaxies would in fact obey the secondary
selection criteria of any given survey (for example, the colour
criteria in Lyman-break selected samples). We defer these sorts of
more detailed comparisons to future work. Figure~\ref{fig:lfzuv} shows
the rest-frame $\sim1500$ \AA~LFs from $z=0.5$ to $z=5$. We see that
the models produce more than enough UV luminous galaxies at all
redshifts, requiring some extinction even at $z\sim5$. However, the
model with fixed dust parameters (normalized at $z=0$) systematically
underproduces UV-luminous galaxies by a larger and larger factor as
redshift increases. This has been found before by other studies using
semi-analytic models \citep{lofaro:09,guo-white:09}. Moreover, as
already mentioned, there is direct observational evidence that high
redshift galaxies may be less extinguished than their low redshift
counterparts \citep{reddy:10}. With the simple evolving dust model, we
obtain fairly good agreement over all redshifts, although we somewhat
overproduce low-luminosity galaxies at high redshift. Obviously, we
could have adopted a more complex dust model with luminosity dependent
evolution in the dust parameters to get a better overall
fit. Alternatively, this could be an indication that low-mass galaxies
are forming too early in the models \citep{fontanot:09}.

\begin{figure*} 
\begin{center}
\includegraphics[width=6.5in]{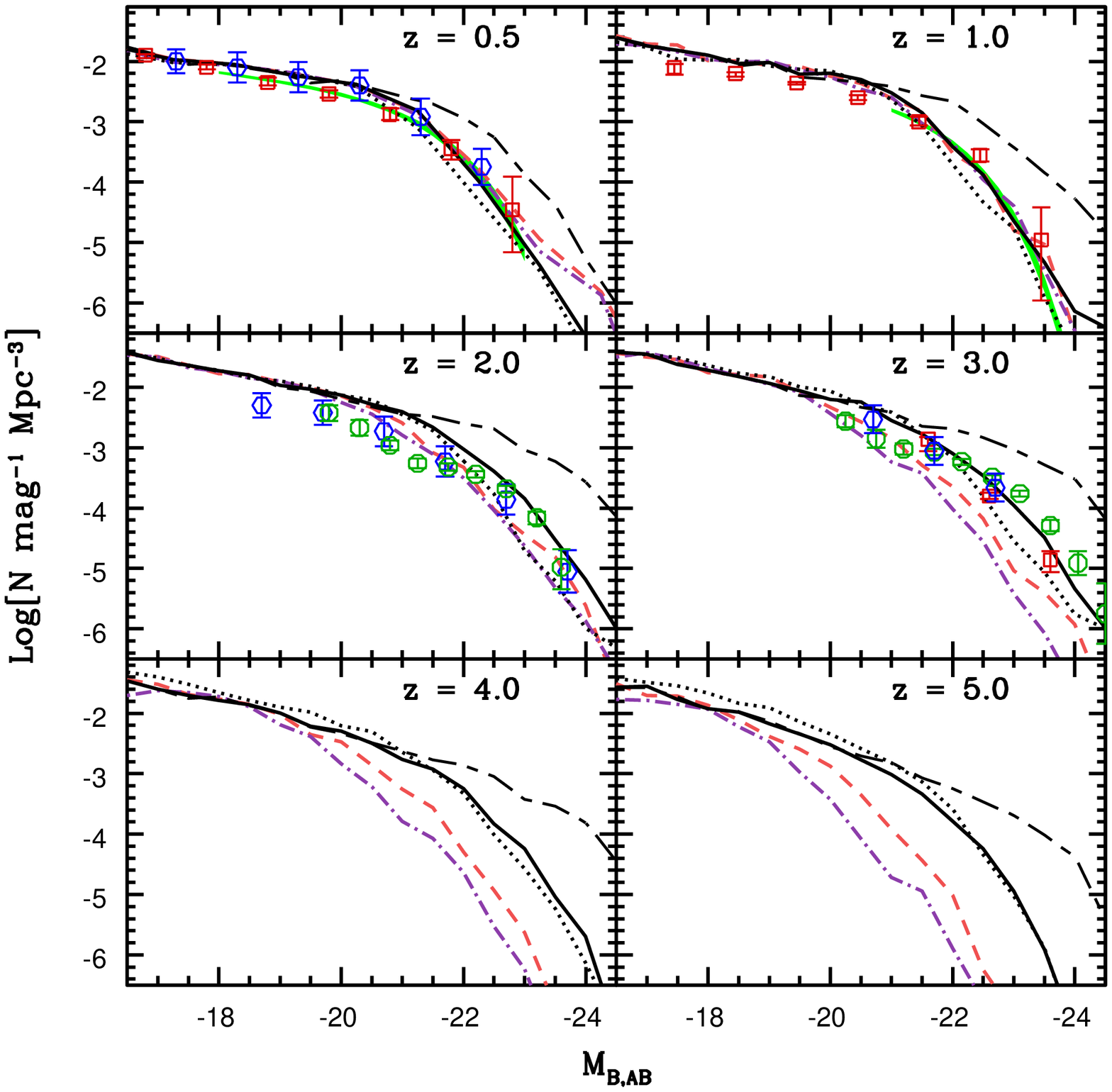}
\end{center}
\caption{Rest frame luminosity functions in the B-band at several
  redshifts.  Line types are the same as in Figure~\ref{fig:lfzuv}.
  Red squares are data from \citet{salimbeni08}, and open blue hexes
  are from the study of \citet{giallongo05}.  The green shaded regions
  in the top two panels are the Schechter fits to DEEP survey
  luminosity functions \citep{faber07}, with 1$\sigma$ errors.  Green
  circles in the two middle panels are from \citet{marchesini07}.
\label{fig:lfz_b}}
\end{figure*}
Figure~\ref{fig:lfz_b} shows a similar comparison in the rest
B-band. We see a similar discrepancy with our non-evolving dust model
to that observed in the UV: we need to adopt lower extinctions in high
redshift galaxies to reproduce the number of luminous galaxies. The
discrepancy is smaller, and sets in at higher redshift, but is still
pronounced by $z\sim3$.  The combination of the rest UV and optical
observations are what forced us to adopt evolving parameters in
\emph{both} components of our two-component model (the cirrus and the
birthclouds).

\begin{figure*} 
\begin{center}
\includegraphics[width=6.5in]{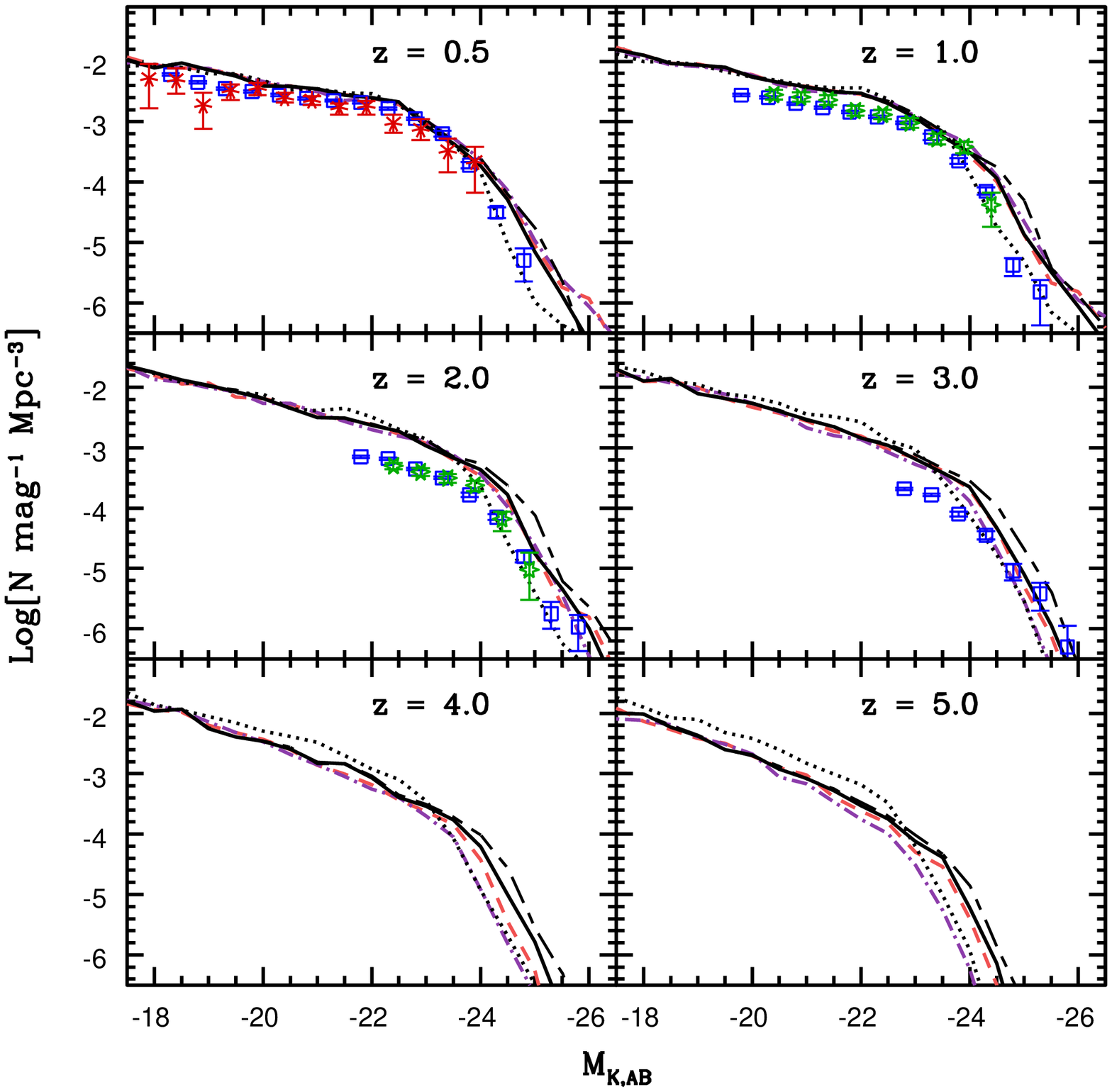}
\end{center}
\caption{Rest frame luminosity functions in the K band at several
  redshifts.  Line types are the same as in Figure \ref{fig:lfzuv}.
  Blue squares are observations from \citet{cirasuolo10}, red stars
  at $z=0.5$ are from \citet{pozzetti03}, and open green stars at redshifts
  1 and 2 are from \citet{caputi06}.  Note that all observations have been
  converted to the AB magnitude system.
\label{fig:lfz_k}}
\end{figure*}

Next, in Figure~\ref{fig:lfz_k} we show the comparison with the
rest-frame K-band luminosity functions in the same redshift
bins. Here, interestingly, the models over-predict the number of
galaxies at high redshift, particularly below the knee in the
luminosity function. This is consistent with the findings of
\citet{fontanot:09}, who showed that three independently developed
semi-analytic models all overproduce low-mass galaxies at high
redshift (see also \citealt{marchesini:09}). The good agreement of our
models at the brightest K-band magnitudes also indicates that,
contrary to the findings of \citet{henriques:10}, there is no need to
invoke an enhanced contribution to the Near-IR due to the Thermally
Pulsating Asymptotic Giant Branch (TP-AGB) stellar phase, as in
e.g. the stellar population models of \citet{maraston:05}.

\begin{figure*} 
\begin{center}
\includegraphics[width=6.5in]{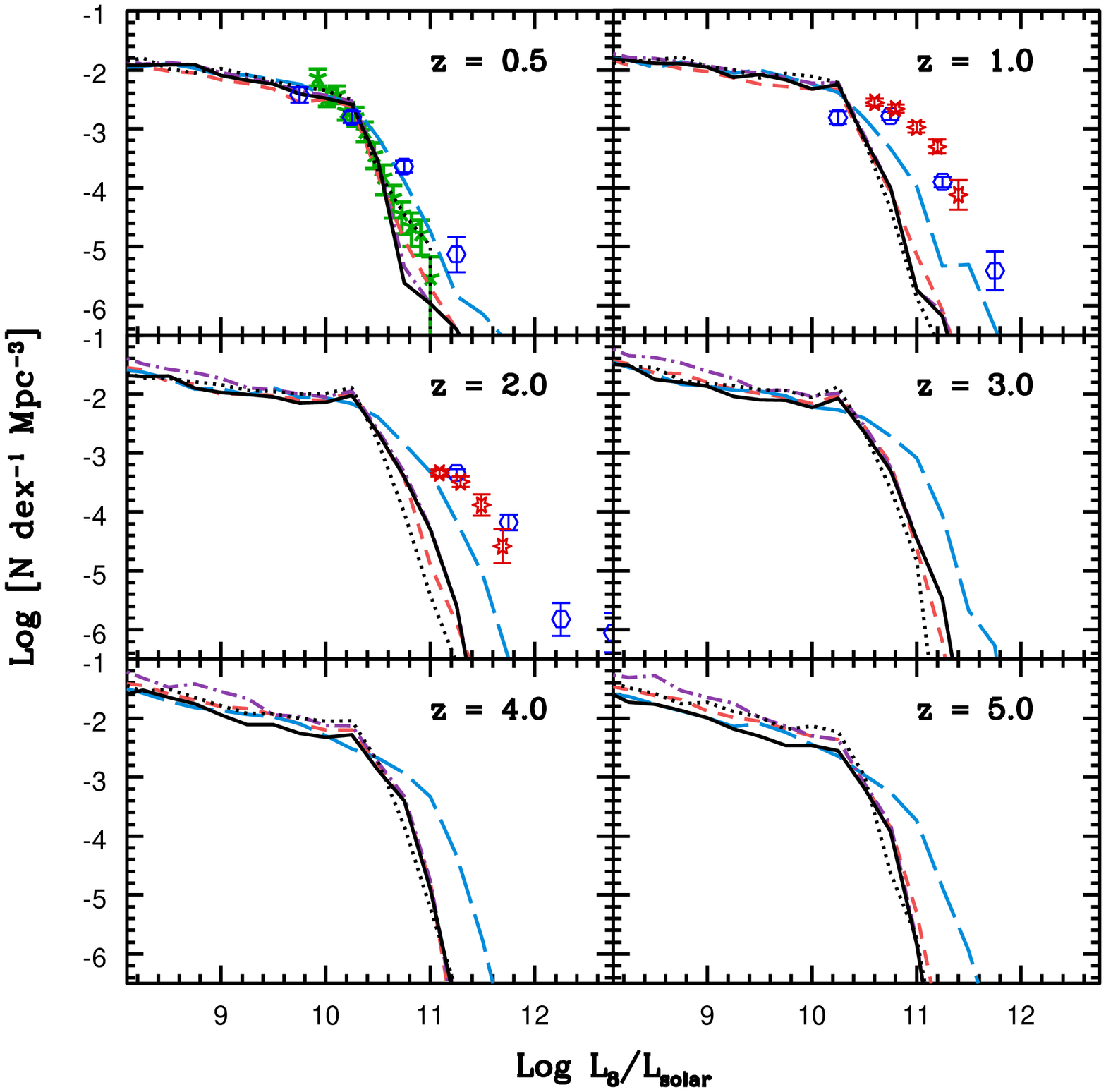}
\end{center}
\caption{Rest frame luminosity functions in the IRAC 8 \mum\ band at
  several redshifts.  Long-dashed blue lines show the predictions of
  the model with the DGS99 dust emission
  templates.  Other lines show results with the IR templates of R09:
  solid black is our fiducial WMAP5 model using our evolving composite
  dust recipe, dashed-dotted purple shows the model with fixed dust
  parameters, dashed red show the single component dust (Calzetti)
  model, and dotted black is the C-\LCDM\ model.  Green stars
  are data from \citet{dai09}.  Blue hexagons show the data from
  \citet{rodighiero10}, and red stars are measurements from
  \citet{caputi07}.
\label{fig:lfz_8mum}}
\end{figure*}

\begin{figure*} 
\begin{center}
\includegraphics[width=6.5in]{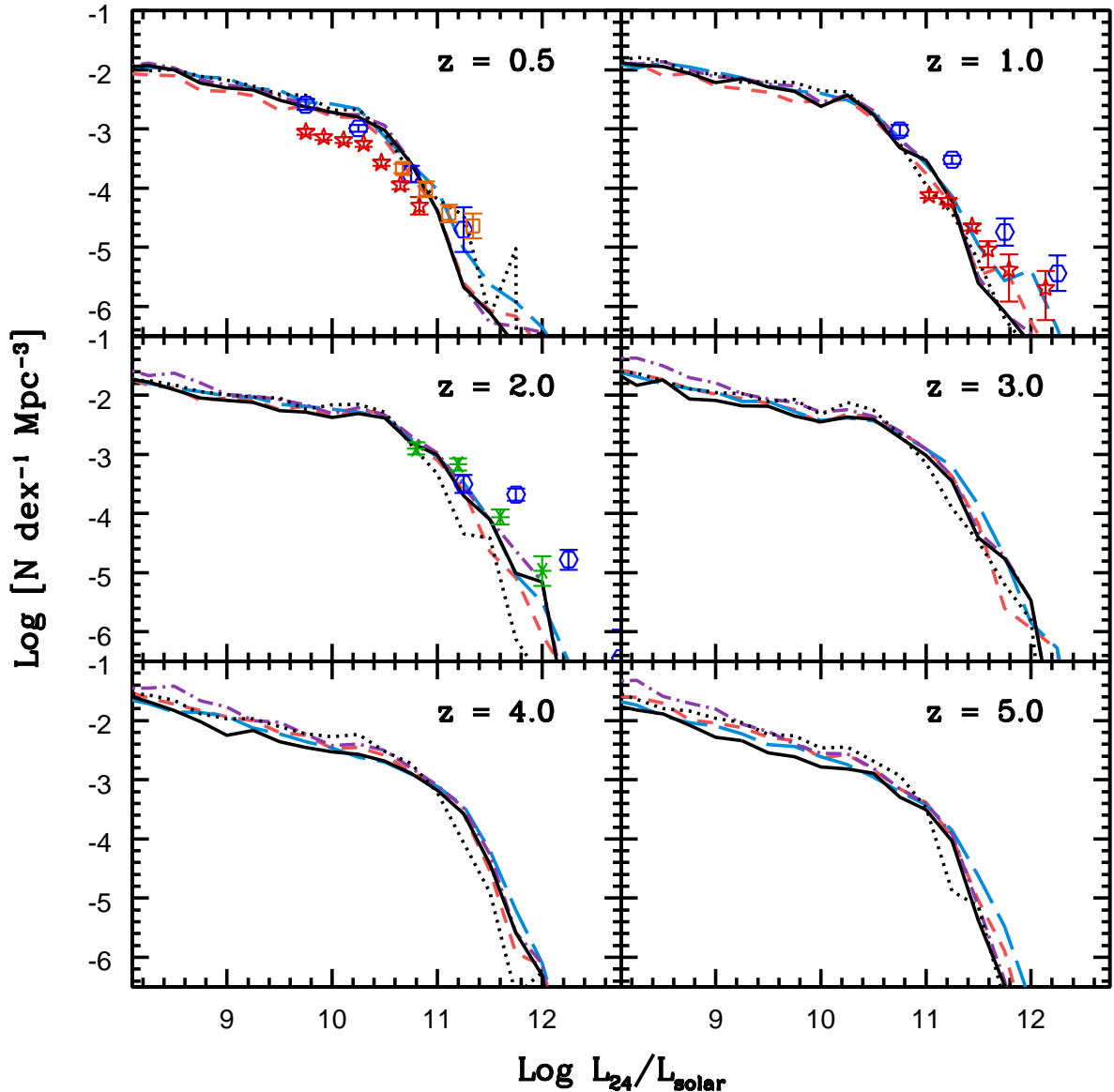}
\end{center}
\caption{Luminosity functions in the rest-frame MIPS 24 \mum\ band.
  Line types are the same as in Figure \ref{fig:lfz_8mum}.  Blue
  hexagons show the observational data from \citet{rodighiero10}, and
  red stars are from \citet{babbedge06}.  Green asterisks are from
  \citet{magnelli11}.  Orange squares are from \citet{rujopakarn10};
  we have interpolated these points from data presented for two
  different redshift bins.
\label{fig:lfz_24mum}}
\end{figure*}

\begin{figure*} 
\begin{center}
\includegraphics[width=6.5in]{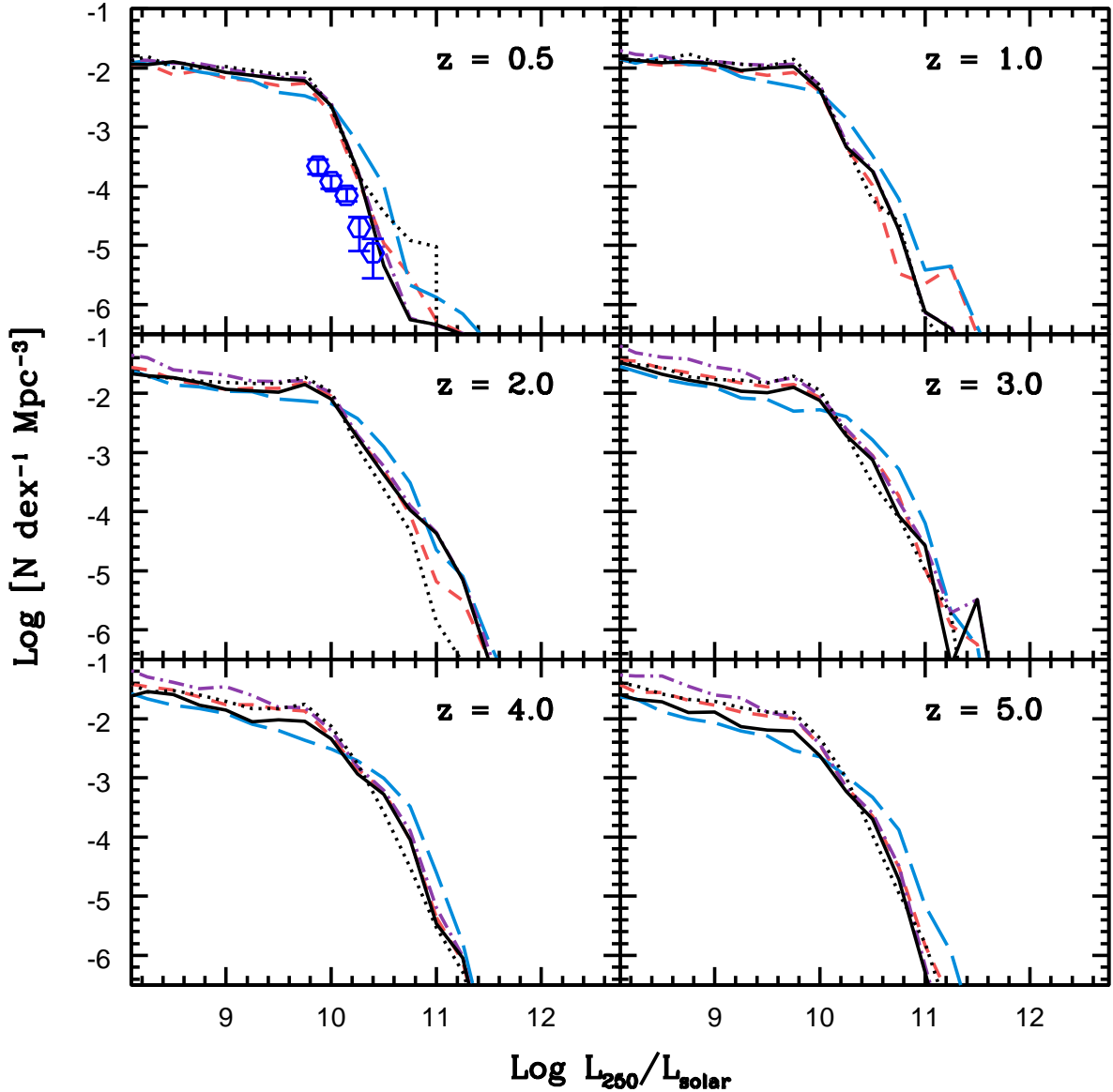}
\end{center}
\caption{The rest-frame luminosity functions at 250 \mum.  Line types
  are the same as in Figure~\ref{fig:lfz_8mum}, and results from early
  Herschel observations \citep{dye10} at z $=0.5$ are shown by the
  blue hexagons.
\label{fig:lfz_250mum}}
\end{figure*}

\begin{figure*} 
\begin{center}
\includegraphics[width=6.5in]{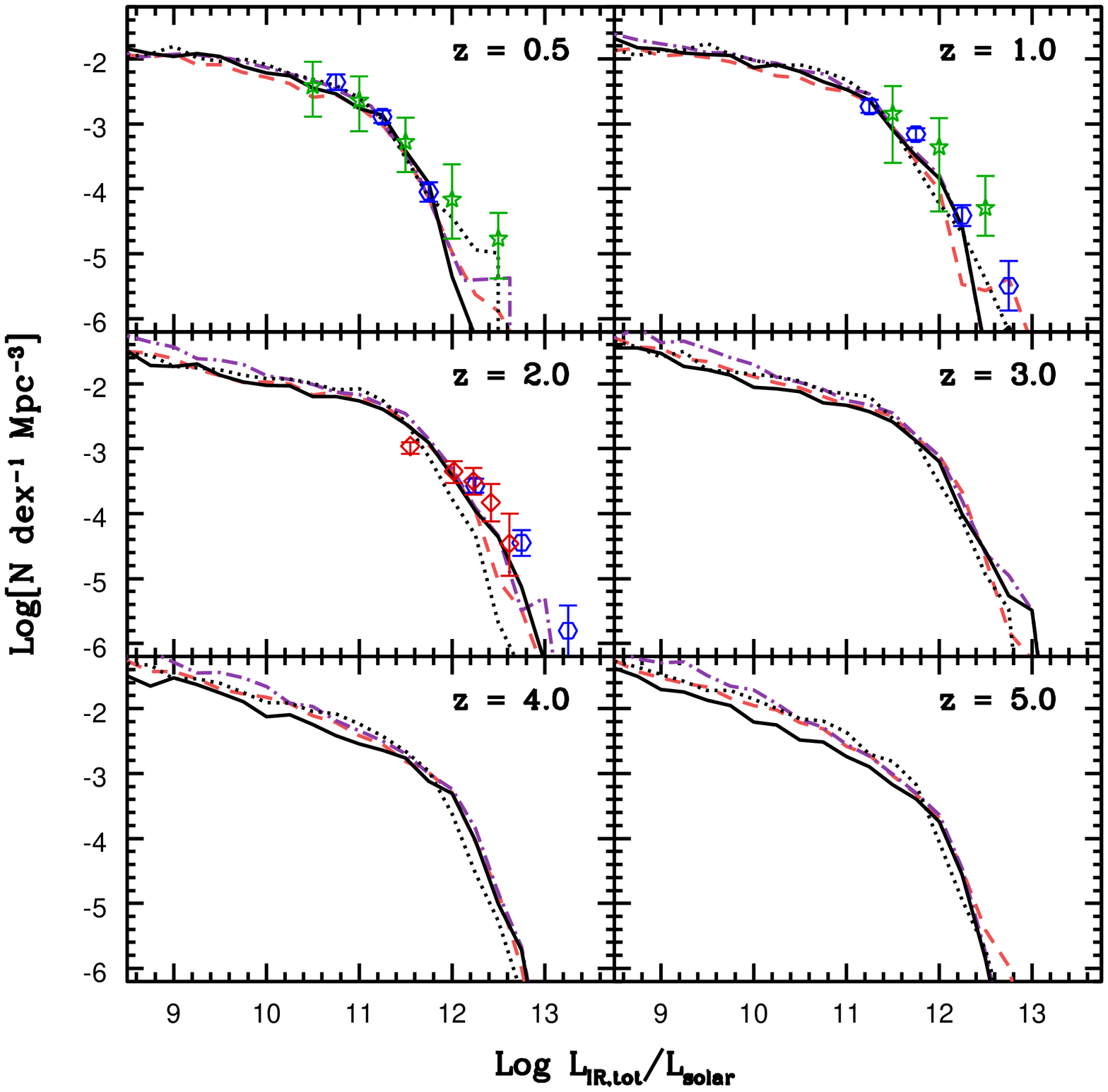}
\end{center}
\caption{Luminosity functions for the total IR emission.  Line types
  are the same an in Figure~\ref{fig:lfz_8mum}, however as the total
  IR predictions are insensitive to the templates used, the model with
  the DGS99 templates is not shown.  Blue hexagons are the data from
  \citet{rodighiero10}, green stars are from \citet{lefloch05}, and
  red diamonds are from \citet{caputi07}.
\label{fig:lfz_irbolo}}
\end{figure*}

We now move from bands that are dominated by light directly emitted by
stars to light that has been reprocessed by
dust. Figure~\ref{fig:lfz_8mum} shows the LF in the Spitzer IRAC rest
8 \mum\ band. Our models produce about the right number of galaxies
around the knee in the LF, but underproduce luminous galaxies,
especially at high redshift ($z\sim 1$--2). As one can see from
Figure~\ref{fig:irtemp}, this portion of the spectrum is very complex,
and highly dependent on whether the spectrum is dominated by emission
from PAHs, or absorption features such as the deep Silicate absorption
seen at $\sim 9 \mu$m. In addition, this part of the spectrum is
particularly subject to contamination by obscured AGN. If there are a
significant number of buried AGN at $z\sim2$, as has been suggested
\citep{daddi:07}, this could at least partially account for the
discrepancy between the observations and our model. However, it would
also be unsurprising if our simple unevolving template approach were
simply inadequate to accurately model the very complex physics that
determines the 8 \mum\ flux in galaxies. Indeed, we see a large
difference between the predictions using the R09 templates and those
using the DGS99 dust emission templates. Very similar remarks apply to
the comparison between the Spitzer MIPS 24 $\mu$m LFs in Figure
\ref{fig:lfz_24mum}, although here the models agree reasonably well
with the observations at $z=0.5$ and $z=1$, and show a somewhat milder
discrepancy at $z=2$. The differences in the predictions with the two
different dust emission templates are also considerably smaller than
at 8 \mum.

Figure~\ref{fig:lfz_250mum} shows the LF at rest 250 \mum, compared
with early results from Herschel. At $z=0.5$, the only
redshift where the LF at this wavelength has been measured to date,
the models overpredict the number of luminous galaxies. The last
luminosity function we examine is that in Figure \ref{fig:lfz_irbolo} for the total integrated IR
from 8--$1000\,\, \mu$m, estimated from multi-wavelength
observations. Here we see fairly good agreement between the models and
the observational estimates, although with a small deficit of
luminous galaxies, particularly at $z\sim2$. However, considering that
the true errors on the observations are probably considerably larger
than the error bars shown, this level of agreement is encouraging. It
is also encouraging that the different model variants produce very
similar results for this quantity.

\begin{figure*} 
\begin{center}
\includegraphics[width=6.5in]{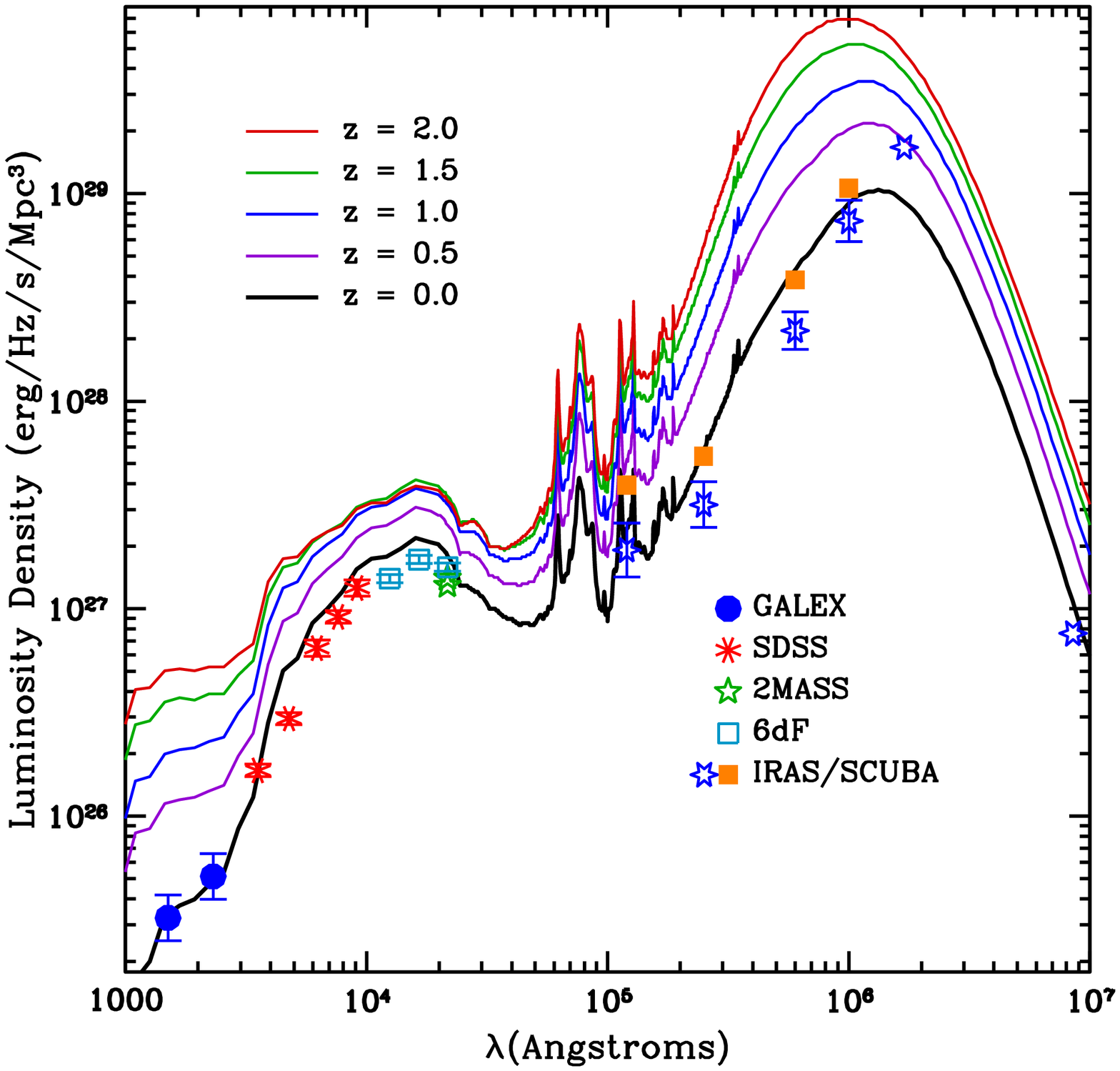}
\end{center}
\caption{Integrated luminosity density as a function of wavelength in
  our WMAP5 fiducial model, shown at various slices in redshift.  All
  data shown are measurements in the local universe.  Measurements are
  from GALEX (blue circle), SDSS (red stars; \citealp{md&prada09}),
  6dF (cyan squares; \citealp{jones06}), and 2MASS (green stars;
  \citealp{cole01} and \citealp{bell03}).  In the mid- and far-IR, the
  orange squares \citep{soifer91} and blue stars \citep{takeuchi01}
  show observations from IRAS and SCUBA. }
 
\label{fig:lumdens_ev1}
\end{figure*}

\begin{figure*} 
\begin{center}
\includegraphics[width=6.5in]{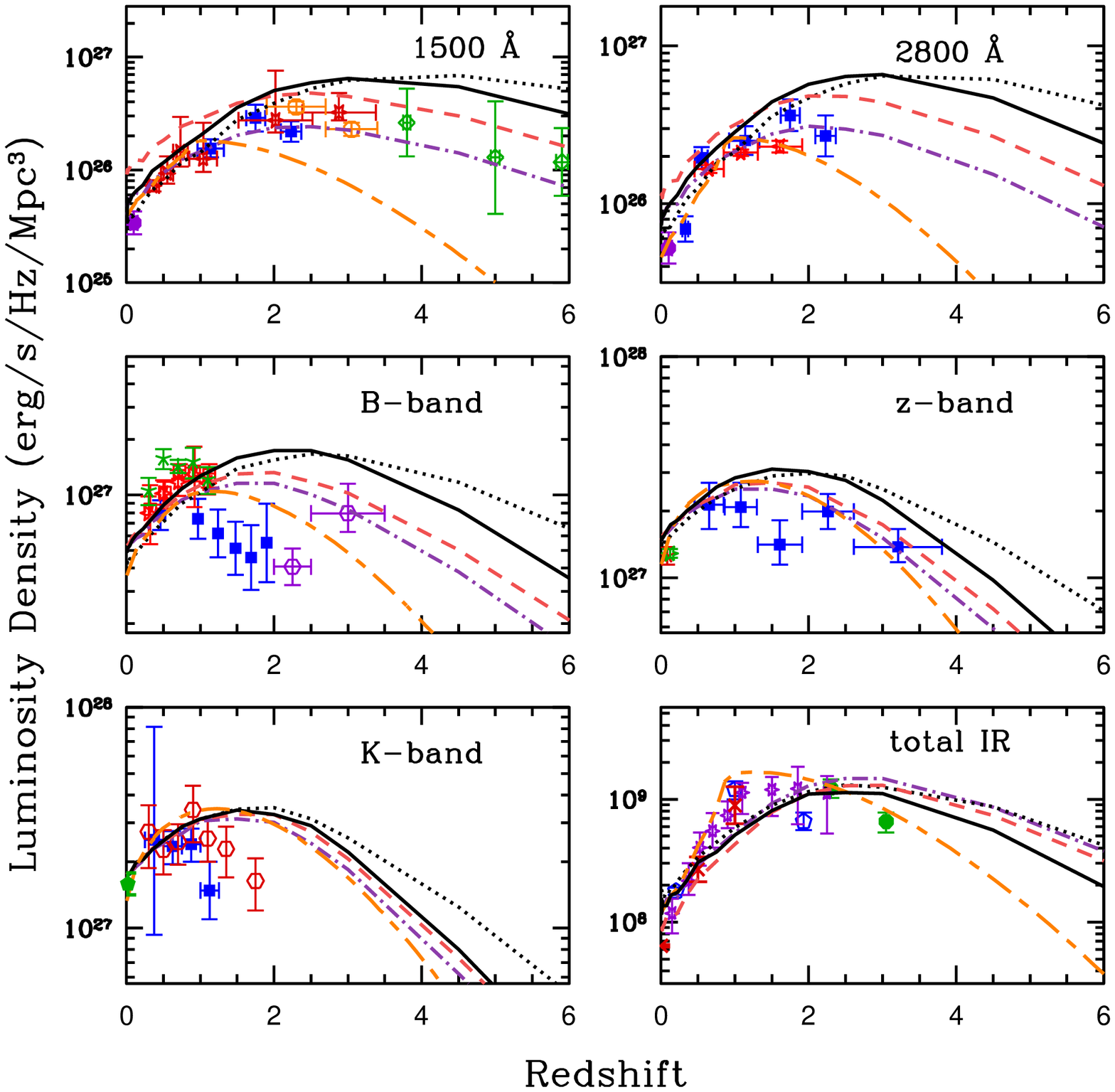}
\end{center}
\caption{Luminosity density as a function of redshift for various
  wavelengths, as well as for the total IR luminosity (8-1000 \mum).
  Line types are the same as those discussed in previous plots; see
  captions for Figures \ref{fig:lfzuv} and \ref{fig:lfz_8mum} and
  Table \ref{tab:models}.  Additionally, orange lines with alternate
  short and long dashes are the predictions of the empirical model of
  \citet{dominguez:11}, for comparison with our work.  
Observational data are as follows: {\bf 1500 \AA:} Blue squares are
from \citet{dahlen07}, red stars are from \citet{schiminovich05},
green stars are from \citet{bouwens07}, and orange circles are from
\citet{reddy08}.  The solid purple circle is a local measurement with
GALEX by \citet{wyder05}. {\bf 2800 \AA:} Blue squares and the purple
circle are again from \citet{dahlen07} and \citet{wyder05},
respectively.  Red stars are from \citet{gabasch06}.  {\bf B-band:}
Blue squares are from \citet{dahlen05}, and are incomplete at higher
redshifts.  DEEP and COMBO-17 data from \citet{faber07} are shown as
red stars and open red squares, respectively (these are very similar
and difficult to differentiate here).  Other data shown are from
\citet{wolf03} (green star) and \citet{marchesini07} (open purple
hexes).  {\bf z-band:} Local measurements are provided by
\citet{md&prada09} (red) and \citet{blanton03} (green).  Blue squares
are from \citet{gabasch06}.  {\bf K-band:} The local determination is
from \citet{kochanek01}.  High redshift data are from \citet{barro09}
(blue squares) and \citet{arnouts07} (open red hexagons).  {\bf Total
  IR Luminosity:} observational estimates of the IR emissivity are
from \citet{caputi07} (open blue pentagons), \citet{reddy08} (green
circles), \citet{rodighiero10} (purple stars), and \citet{lefloch05}
(red crosses).
\label{fig:lumdens_ev2}}
\end{figure*}

Figure~\ref{fig:lumdens_ev1}
shows the total emissivity from all galaxies as a function of
wavelength in our fiducial model, at various redshifts. One can see a
slight shift in the wavelength of the peak of the dust emission, as
well as the ratio of warm to cold dust with redshift. Because we have
assumed non-evolving dust emission templates in this work, this is
entirely due to the changing mix of galaxies of different total IR
luminosities with time (i.e. the declining contribution of very IR
luminous galaxies with decreasing redshift). 
Figure~\ref{fig:lumdens_ev2} shows the integrated luminosity density
as a function of redshift in the far and near-UV, the B-, z-, and
K-band, and the total IR, as a function of redshift for all the models
presented in this work. The model predictions are compared with
observational estimates obtained by integrating observed LF's (see
figure caption for references). Such observational estimates in the UV
and IR bands are often used to estimate the global star formation rate
density (e.g. as shown in Figure~\ref{fig:sfhist}). It is interesting
that while our fiducial model is in good agreement with the
observations in the Far-UV over the whole redshift range, it is
somewhat low compared with the total-IR luminosity at $z\sim 1$--2. As
already discussed, this could be due to the total IR having been
overestimated, or could indicate that the star formation rate in the
models is too low. Also interesting to note is that the UV and IR
luminosity densities peak at a higher redshift ($z\sim 3$) than the z-
or K-band, because the latter arise from older stellar populations.

\subsection{Galaxy Counts}

\begin{figure*} 
\begin{center}
\includegraphics[width=6.5in]{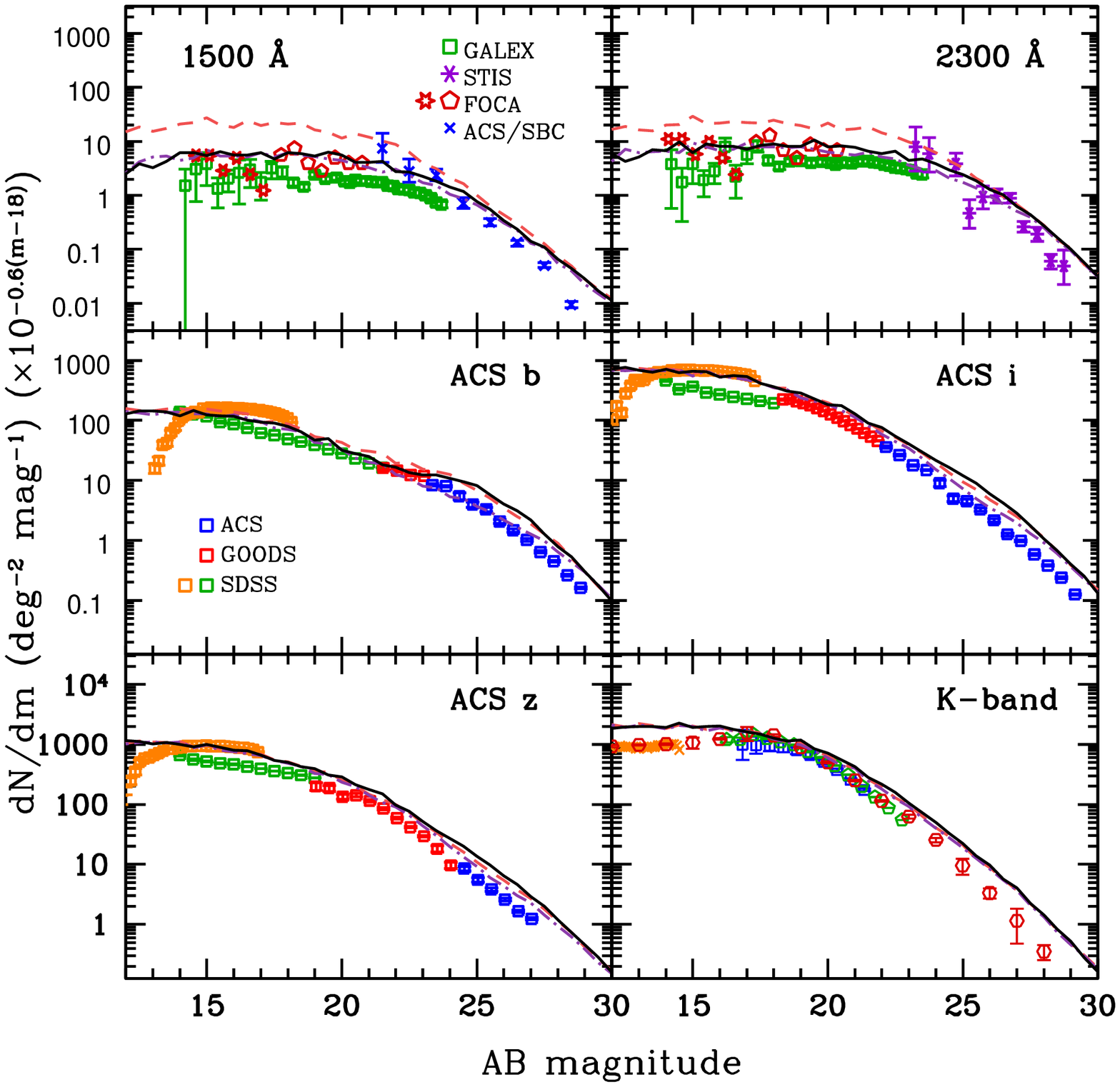}
\end{center}
\caption{Number counts in the GALEX UV bands and the four HST ACS
  bands.  Line types are the same as in Figure \ref{fig:lfz_8mum};
  note that some models do not deviate significantly from the fiducial
  WMAP5+evolving dust model (solid black line) and are therefore not
  visible.  Note that results here have been rescaled to a Euclidean
  geometry.  In the UV bands, data are from {\it GALEX} \citep[][green
    squares]{xu05}, STIS on {\it HST} \citep[][purple
    asterisks]{gardner00}, and the balloon-borne FOCA experiment
  \citep[][red stars and open pentagons
    respectively]{iglesias04,milliard92}.  The FOCA points have been
  converted to the GALEX bands using the method described in
  \citet{xu05}.  Blue crosses are from HST ACS/SBC observations of
  multiple fields in GOODS-N and -S \citep{voyer11}.  In the ACS
  bands, red, blue and green squares are from the compilation by
  \citet{dolch11}, which includes data from the Hubble Ultra-Deep
  Field.  Additional data in orange from SDSS-DR6 are from
  \citet{md&prada09}.  In the K-band, we show data from 6dF (orange
  crosses, \citealp{jones06}), from \citet[][open red
    hexagons]{keenan10}, and from \citet[][blue squares]{barro09}, and
  \citet[][green pentagons]{mccracken10}.  All observational data have
  been converted to AB magnitudes.
\label{fig:countsacs}}
\end{figure*}

\begin{figure*} 
\begin{center}
\includegraphics[width=6.5in]{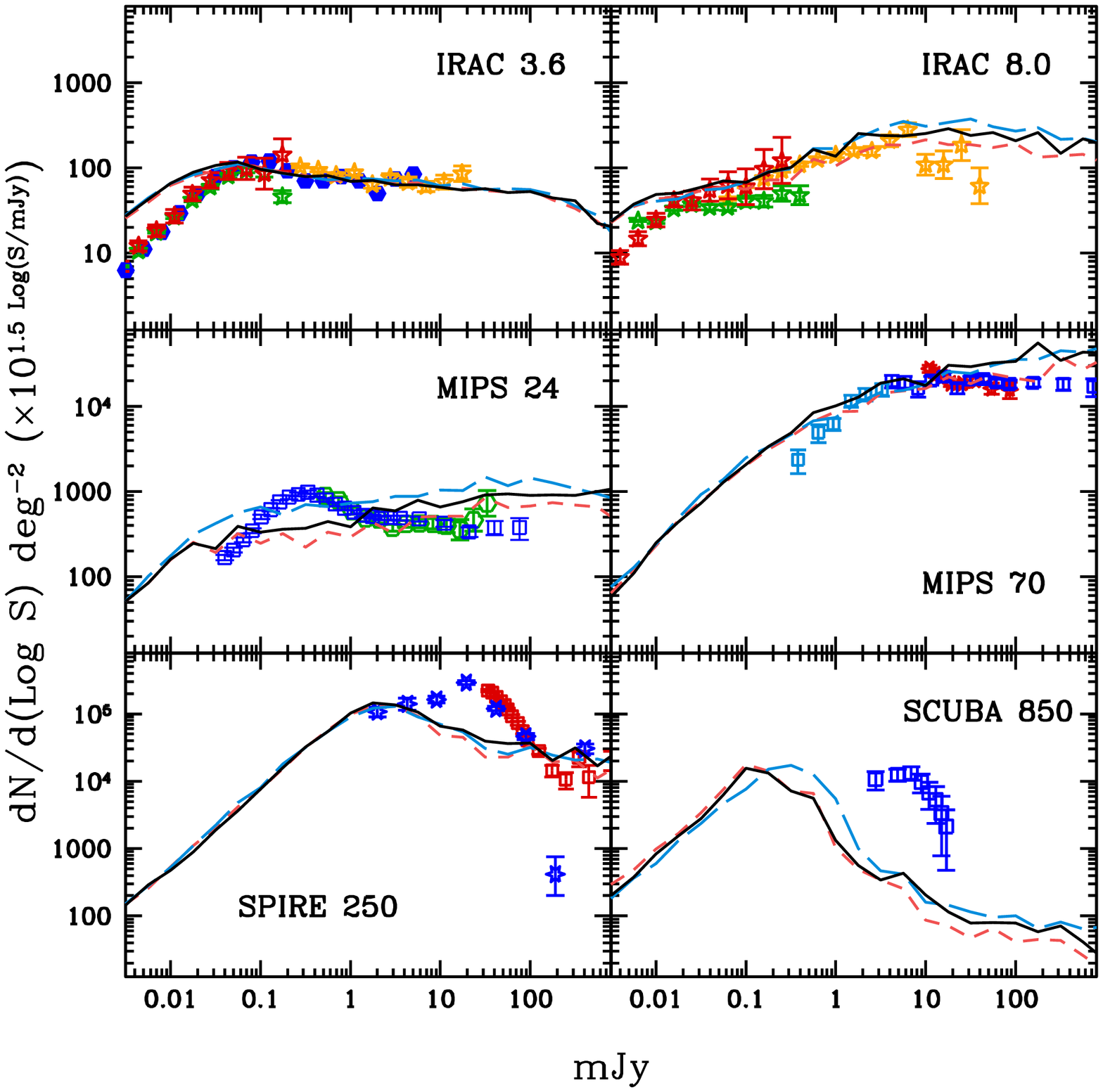}
\end{center}
\caption{Number counts from four Spitzer (IRAC and MIPS) infrared
  bands, as well as Herschel 250 \mum\ and SCUBA 850 \mum.  Line types
  are the same as in Figure \ref{fig:lfz_8mum}; for clarity models
  similar to the fiducial model are not shown.  Results are scaled to
  a Euclidean geometry.  Solid blue circles in the 3.6 IRAC band are
  from \citet{sanders07}; all other points in the IRAC 3.6 and 8.0
  bands are from \citet{fazio04}.  The MIPS data at 24 \mum\ shown here
  are the S-COSMOS `Extragalactic Wide' points from \citet{sanders07}
  (green hexes), and the Wide and Deep Legacy Survey points from
  \citet{bethermin10} (blue squares).  At 70 \mum\ data shown are the
  normal (blue squares) and stacked (cyan squares) measurements from
  \citet{bethermin10}, while red stars are from \citet{frayer06}.
  Herschel data at 250 \mum\ are from \citet[][red squares]{clements10}
  and \citet[][blue stars]{glenn10}; the latter is from the spline
  model with FIRAS priors.  We show data from the SCUBA SHADES survey
  \citep{coppin06} at 850 $\mu$m in the lower-right panel.
\label{fig:countsir}}
\end{figure*}

Figures~\ref{fig:countsacs} and \ref{fig:countsir} show galaxy number
counts from the UV through the sub-mm. This is an important
cross-check on the luminosity function comparison as many
high-redshift samples do not have spectroscopic redshifts available
particularly for the fainter objects. Our fiducial model provides
quite good agreement with the data at all UV and optical
wavelengths. There is a factor of 2--3 excess of faint galaxies in the
models in the optical and NIR, which corresponds to the excess of
faint galaxies at $z\sim 1$--2 seen in the luminosity function
comparison. The agreement at 3.6 and 8 $\mu$m is also quite good,
except below $\sim 30 \mu$Jy, where the models overpredict faint
galaxies. The agreement is also good for MIPS 24 and 70 \mum, except
for 24 \mum\ sources between 100 $\mu$Jy to 1 mJy.  The models with
the R09 templates underproduce galaxies in this range, while the model
with the DGS99 templates performs better. This is because of the much
lower fluxes in the 10-20 \mum\ range in the R09 templates relative to
the DGS99 templates.  None of the models reproduce the ``bump'' seen
in the observed 24 \mum\ counts between $\sim 0.1$ and 1 mJy. The
agreement with the 70 \mum\ counts is fairly good, but the observed
counts do not go as deep.

Moving towards longer wavelengths, the agreement becomes increasingly
poor. Our models underpredict galaxies fainter than $\sim 80 \mu$Jy at
250 \mum\ by a factor of 4--5, and underproduce sub-mm galaxies at 850
\mum\ by more than an order of magnitude. It is well-known that
CDM-based models in general have difficulty reproducing the observed
population of sub-mm galaxies
\citep[e.g.][]{devriendt:00,baugh:05}. We discuss different possible
interpretations of these results in Section~\ref{sec:discussion}, and
we present more detailed predictions for galaxies in the Herschel PACS
and SPIRE wavebands in \citet{niemi:12}.
%(Niemi et al., in prep.).

\subsection{The Extragalactic Background Light}

\begin{figure*} 
\begin{center}
\includegraphics[width=5.0in]{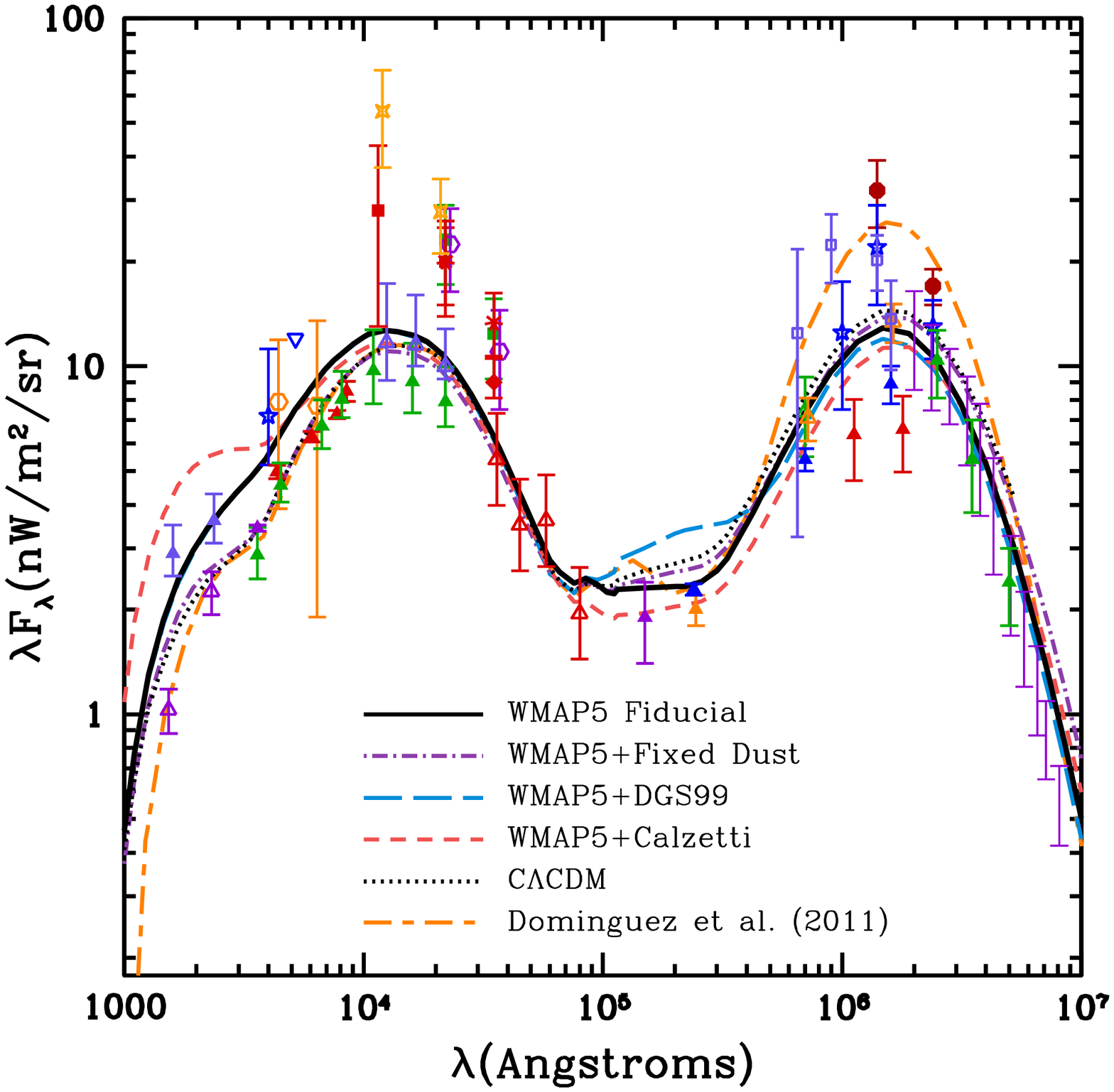}
\end{center}
\caption{The predicted integrated Extragalactic Background Light
  spectrum, compared with experimental constraints.  Line types follow
  the convention of earlier plots.
{\bf Data:} Upward pointing arrows indicate lower bounds from number
counts; other symbols are results from direct detection experiments.
Note that some points have been shifted slightly for clarity.  {\bf
  Lower limits:} The blue-violet triangles are results from Hubble and
STIS \citep{gardner00}, while the purple open triangles are from GALEX
\citep{xu05}.  The solid green and red triangles are from the Hubble
Deep Field \citep{madau00} and Ultra Deep Field \citep{dolch11}
respectively, combined with ground based-data, and the solid purple
triangle is from a measurement by the Large Binocular Camera
\citep{grazian09}.  In the near-IR J, H, and K bands, open violet
stars are the limits from \citet{keenan10}.  Open red triangles are
from IRAC on Spitzer \citep{fazio04}, and the purple triangle at 15
$\mu$m is from ISOCAM \citep{hopwood10} on ISO.  
The lower limits from MIPS at 24, 70, and 160 $\mu$m on Spitzer are
from \citet{bethermin10}, \citet{chary04}, \citet{frayer06}, and
\citet{dole06} (solid blue, solid gold, open gold, and open green,
respectively).
Lower limits from Herschel number counts \citep{berta10} are shown as
solid red triangles.  In the submillimeter, limits are presented from
the BLAST experiment \citep{devlin09}. {\bf Direct Detection:} In the
optical, orange hexagons are based on data from the Pioneer 10/11
Imaging Photopolarimeter \citep{matsuoka11}.  The blue star is a
determination from \citet{mattila11}, and the triangle at 520 nm is an
upper limit from the same.  In the near-IR, the points at 1.25, 2.2,
and 3.5 $\mu$m are based upon DIRBE data with foreground subtraction:
\citet[][dark red squares]{wright01}, \citet[][orange
  crosses]{cambresy01}, \citet[][red diamond]{levenson08},
\citet[][purple open hexes]{gorjian00}, \citet[][green
  square]{wright&reese00}, and \citet[][red asterisks]{levenson07}.
In the far-IR, direct detection measurements are shown from DIRBE
\citep[][blue stars and solid red circles]{wright04,schlegel98}, and
FIRAS \citep[][purple bars]{fixsen98}. Blue-violet open squares are
from direct photometry with the AKARI satellite \citep{matsuura10}.
\label{fig:ebl}}
\end{figure*}

One of the major goals of this work is to predict the integrated
Extragalactic Background Light (EBL). The EBL consists of light
emitted at all epochs, modified by redshifting and dilution due to the
expansion of the universe. 
Because the EBL consists of light emitted by all types of sources
  at all cosmological epochs, it forms a unique record of the history
  of photon production in the universe.  A detailed measurement of the
  EBL flux can potentially inform us about the existence or
  nonexistence of photon sources beyond those that can be resolved in
  galaxy surveys, and its SED encodes the details of the redshifts and
  spectral characteristics of these sources.  As discussed in our
  companion paper (GSPD), the EBL also affects the propagation of
  gamma rays in the GeV and TeV bands. 
The EBL at a redshift $z_0$
and frequency $\nu_0$ in proper coordinates can be computed in our
model as the following integral over emissivities from our predicted
galaxy population:
\begin{equation}
J(\nu_0,z_0)=\frac{1}{4\pi} \int^{\infty}_{z_0} \frac{dl}{dz} \frac{(1+z_0)^3}{(1+z)^3}\epsilon (\nu,z) dz,
\end{equation}
where $\epsilon(\nu,z)$ is the galaxy emissivity at redshift $z$ and frequency $\nu=\nu_0(1+z)/(1+z_0)$, and $dl/dz$ is the cosmological line element, defined as
\begin{equation}
\frac{dl}{dz}=\frac{c}{(1+z)H_0} \frac{1}{\sqrt{\Omega_m(1+z)^3+\Omega_\Lambda}}
\label{eq:cosline}
\end{equation}
for a flat $\Lambda$CDM universe \citep{peebles93}.  We assume here
that the EBL photons evolve passively after leaving their source
galaxies and are not affected by any further interactions except for
cosmological redshifting.  This is an acceptable approximation for
photons at energies below the Rydberg energy of 13.61 eV. 

 We note that our estimate of the EBL does not include the
  contribution from light radiated by AGN. However, previous studies
  have shown that this contribution should be less than $\sim 10-20$\%
  in the mid-IR, and smaller (a few percent) at other wavelengths
  \citep[see][and references therein]{dominguez:11}.

Observational estimates of the EBL are obtained either by direct
detection or by integration of galaxy counts.  Direct detection is
complicated by foreground emission from our own galaxy and reflected
zodiacal light from our sun, which are much brighter than the EBL
across most of the optical and IR spectrum
\citep{hauser&dwek01}. Integrated galaxy counts provide a firm lower
limit, but there has been considerable debate about how much light
these estimates might be missing because of undetected, low surface
brightness galaxies or the faint extended wings of detected galaxies,
which can be difficult to disentangle from the
background. Figure~\ref{fig:ebl} shows the predictions of our five
model variants along with a compilation of recent observational
estimates of the EBL at various wavelengths from both methods. There
is a significant gap between the direct detection measurements and the
integrated counts, with the former always lying higher than the
latter, indicating that there must still be biases or systematic
sources of error in one or both methods. Unsurprisingly, since we have
already seen that our models reproduce the galaxy counts at most
wavelengths, our model predictions all lie close to the estimates from
integrated counts. It is noteworthy that although our models, if
anything, seem to \emph{over}-produce faint galaxies, the integrated
EBL predictions are nowhere nearly as high as the direct detection
estimates. The model with the one-component (Calzetti) attenuation law
over-produces the far-UV EBL and is low in the mid-IR. This is because
in the two-component (modified Charlot-Fall) dust model, young stars
are enshrouded in dense birth clouds with higher optical depth. The
largest difference between the models with different dust emission
templates is seen in the mid-IR, with the DGS99 models predicting a
higher EBL in the mid-IR than the R09 models. The two models bracket
the errorbars on the existing observations. Hopefully new observations
will tighten up these constraints. The peak of the EBL at $\sim
100-250$ \mum\ is a bit low compared with observational estimates, but
is within the errors. In terms of the overall partition of the EBL in
the UV-NIR vs. NIR-FIR regimes, our models are in good agreement with
the observational constraints (see GSPD).

%=======================
% 4
\section{Discussion}
\label{sec:discussion}
%=======================

\subsection{Comparison with Previous Results}

\citet{guiderdoni:98} was one of the earliest studies using ``forward
evolution'', cosmologically motivated CDM models combined with
modelling of both dust extinction and emission. This work was extended
in \citet{devriendt:00} and subsequent papers by the GALICS
collaboration \citep{hatton:03,blaizot:04}. They coupled a
semi-analytic model with analytic recipes for dust attenuation and
theoretical templates for dust emission, very much in the same spirit
as our work here. The \citet{devriendt:00} models overpredicted the UV
and optical counts, possibly because of their use of a single dust
attenuation relation rather than a two-component model like our
modified Charlot-Fall model. They also found that their standard model
failed to reproduce enough sub-mm galaxies at 850 \mum, but were able
to fit the sub-mm counts by introducing by hand a population of
heavily dust extinguished starburst galaxies at high-redshift.

The Durham group has coupled their semi-analytic model of galaxy
formation \citep{cafnz:94,cole:00} with the dust models and radiative
transfer code GRASIL, developed by \citet{silva:98}. An important
feature of the GRASIL approach is that the dust emission SED is
computed self-consistently based on the assumed properties of the dust
and the predicted radiation field.  This work is presented in an
extensive series of papers
\citep{granato:00,baugh:05,swinbank:08,lacey:08,lacey:10}. \citet{granato:00}
showed that the fiducial model of \citet{cole:00}, when coupled with
the GRASIL machinery, could reproduce local galaxy luminosity
functions from the UV to the FIR (12-100 \mum). However,
\citet{baugh:05} showed that this model failed to reproduce sufficient
numbers of bright sub-mm galaxies by an order of magnitude or
more. They therefore made several modifications to their model. They
modified the star formation recipes in their model in two ways: 1)
they adopted a quiescent star formation recipe with a constant star
formation timescale, rather than a standard Kennicutt-Schmidt
relation, leading to larger gas reservoirs in high redshift galaxies;
2) they adopted an efficient starburst mode in minor as well as major
mergers. As shown by \citet{spf:01}, this leads to a much larger
population of luminous starburst galaxies at high redshift. In
addition, they adopted a top-heavy stellar Initial Mass Function (IMF)
in bursts, in which the slope of the IMF $dN/d\ln m \propto m^{-x}$ is
$x=0$ over the whole range $0.15 < m < 125$ \msun. In addition to
producing much more UV light per unit mass of stars formed, the
top-heavy IMF also produces much larger amounts of metals and
dust. They found that all of these changes combined were needed to
reproduce the sub-mm counts. The same model simultaneously reproduces
the rest-UV and optical LFs at high redshift.

\citet{lacey:08} argue that the top-heavy IMF is needed in order to
reproduce the evolution of the mid-IR LF as measured by Spitzer, and
the counts in Spitzer bands as well. However, \citet{swinbank:08}
showed that the K-band and IRAC 3-8 \mum\ luminosities of both SMGs and
LBGs in the Lacey-Baugh models are a factor of ten too low (because of
the dearth of low-mass stars). Moreover, the predictions of the models
do not seem to be in good agreement with early results from BLAST and
Herschel at 350--500 \mum\ \citep{lacey:10,clements10}.

Comparison between our results and those of Lacey et al. is
complicated because not only the approach for treating dust is
different, but also many of the other model ingredients are
significantly different. Perhaps most significantly, the published
Lacey-Baugh models do not include ``radio mode'' feedback from AGN,
which is now commonly adopted in semi-analytic models in order to
prevent the formation of grossly over-massive galaxies
\citep[e.g.][]{croton:06,bower:06,s08}. In an attempt to
solve the overcooling problem via other means, \citet{cole:00} adopted
modified hot gas profiles that suppressed early cooling in massive
haloes. This also suppresses the early formation of massive galaxies
\citep{bower:06}. Interestingly, \citet{fontanot:07} use basically the
same approach to modeling dust (GRASIL) within a different
semi-analytic model, and find that they are able to reproduce the
K-band and sub-mm counts with a Salpeter IMF. They ascribe this
success to an improved cooling model. However, their models
overproduce massive galaxies at low redshift ($z<1$). Clearly, more
work is needed in order to develop a model that reproduces all the
available observations.

One should also keep in mind that the GRASIL dust+RT models used in
the work described above assume a simplified, regular geometry
(spheroid+disc). Particularly when considering populations that are
likely to correspond to merger-driven starbursts, such as the SMGs, it
is probably important to capture the complex physics and geometry of
these systems. Some significant progress has been made recently in
this regard, using hydrodynamic simulations of isolated and merging
galaxies in combination with the polychromatic Monte Carlo Radiative
Transfer code SUNRISE
\citep{jonsson:06,jonsson06a,jonsson:10,jonsson&primack10}.
\citet{narayanan:10a} studied the SEDs of merger simulations combined
with SUNRISE RT calculations and identified objects with properties
consistent with both bright and fainter SMGs. \citet{narayanan:10b}
used similar simulations to propose a physical model for the $z\sim2$
DOGs (Dust Obscured Galaxies -- optically faint galaxies identified at
24 \mum), which partially overlap with the SMG population. The SEDs
from these simulations can be combined with predictions of merger
rates from semi-empirical models like those of \citet{hopkins:10}, or
with semi-analytic models, to compute statistics of the population,
such as counts (Hayward et al. in prep).

There is an extensive literature on ``backwards evolution'' models for
galaxy counts and the EBL, which we review in our companion paper
GSPD, but do not discuss here. Recently, empirical and
``semi-empirical'' approaches have been adopted by several authors to
predict the EBL. \citet{younger:11} used the observed stellar mass
function at different redshifts, in combination with a semi-empirical
model of galaxy evolution and template SEDs from
\citet{chary-elbaz:01} to predict the mid to FIR EBL.
\citet[][D11]{dominguez:11} made use of empirical
template SEDs and observed fractions of 25 different galaxy types to
predict the EBL and its evolution. An explicit comparison with the
luminosity density evolution estimated by their approach is shown in
Figure~\ref{fig:lumdens_ev2}. The D11 estimates are anchored to the
observed K-band luminosity functions, which our semi-analytic models
reproduce fairly well, so the predictions are very similar at NIR
wavelengths. At shorter wavelengths (optical and UV), the D11 approach
predicts lower luminosity densities at high redshift than our
SAMs. These differences are due to the use by D11 of SWIRE templates
\citep{polletta:07} from the UV to IR, while in our approach we model
the star formation history and dust attenuation of each galaxy.  In
the Far-IR, D11 estimate a higher and sharper peak at $z\sim$1--2
(again because of the use of different SED templates), in better
agreement with observations, and a steeper decline at higher redshift
$z\gtrsim 2$. Note that the observed K-band luminosity functions that
ground their empirical approach are available only up to $z\sim 4$,
and the results shown at higher redshifts are extrapolations.

\subsection{Summary and Conclusions}

We have presented predictions for the luminosity and flux
distributions of galaxies from the far-UV to the far-IR and over the
bulk of cosmic history ($z = 0$--6). Our predictions are based on
semi-analytic models of galaxy formation, set within the hierarchical
Cold Dark Matter paradigm of structure formation, and including
modeling of gas cooling, star formation, stellar feedback, chemical
enrichment, and AGN feedback. In addition, crucial to the present
study is modeling of the attenuation and re-emission of starlight by
dust in the interstellar medium of galaxies. We use a simple but
physically motivated analytic approach to estimate the dust
attenuation as a function of wavelength. In our fiducial models, based
on the approach proposed by \citet{charlot&fall00}, young stars are
enshrouded in dense ``birth clouds'', while older stellar populations
are embedded within a more diffuse ``cirrus'' component. Stars emerge
from the dense birth clouds as they age. This two-component dust model
results in an effectively age-dependent attenuation relation, such
that younger stars are more extinguished. We find that the
two-component model gives much better agreement with the UV-optical
colours of galaxies than the widely used approach of a fixed
attenuation curve.

We then assume that all light absorbed by dust is re-radiated in the
IR, and use a set of ``template'' SEDs to estimate IR
luminosities. Our current assumption is that the total IR luminosity
of a galaxy determines which template SED to use, based on the
empirical correlation between bolometric or total IR luminosity and
dust temperature in local LIRGS and ULIRGS
\citep{sanders:96}. However, this is certainly too simplistic, and a
goal of our future research is to try to understand and characterize
how the physical parameters of galaxies impact the shape of their FIR
SEDs. One avenue towards this goal is to use detailed dust and
radiative transfer models implemented within hydrodynamic simulations
\citep{jonsson:06,jonsson06a,jonsson:10,narayanan:10a,narayanan:10b}. Another
approach is to use the much richer set of mid- and FIR data that is
becoming available, spanning a broader range of cosmic epoch and
galaxy type, to develop a more complete set of template SEDs
\citep[e.g.][]{chary-pope:10}.

We found that in order to fit the luminosity functions and counts in
the UV and (to a lesser extent) optical, it was necessary to adopt a
dust optical depth normalization that varied with redshift. This has
the net effect that galaxies of a given bolometric luminosity are less
extinguished at high redshift, and could be interpreted as an evolving
dust-to-metal ratio or as a geometric effect (e.g. perhaps the
distribution of gas and dust relative to the stars is different in
high redshift galaxies). This result has also been found by other
groups working with semi-analytic models
\citep{lofaro:09,guo-white:09} and is supported by direct
observational evidence \citep{reddy:10}. However, our current approach
is completely ad hoc (we simply tuned the dust normalization
parameters to match the observed FUV luminosity functions from
$z=0$--5) and it would be desirable to develop better observational
constraints as well as a deeper physical understanding of this
effect. With the evolving dust model, we find very good agreement with
observed far-UV luminosity functions from $z\sim 0$--5 and B-band
luminosity functions from $z\sim 0$--3, and slightly over-predict
galaxies in the rest near-IR (K-band) at high redshift ($z\sim
2$--3). In all UV-NIR bands our models predict an excess of
low-luminosity galaxies, which confirms the excess found by
\citet{fontanot:09} and others based on stellar mass function
comparisons.

It would have been unsurprising if our simple approach for computing
IR luminosities disagreed drastically with more detailed radiative
transfer calculations or with observations. However,
\citet{fontanot:09a} and \citet{fontanot:10} showed that implementing
a recipe similar to the one we adopt here within the MORGANA SAM
produced very similar results to the full radiative transfer
calculations using the GRASIL code of \citet{silva:98}, at least for
global quantities such as luminosity functions and counts. And we have
shown that our models reproduce galaxy counts from the UV to the
mid-IR ($\sim 70$ \mum) remarkably well. A growing discrepancy starts
to emerge at longer wavelengths: our models underproduce intermediate
luminosity ($\sim 30$--100 mJy) sources in the SPIRE 250 \mum\ band by
a factor of 2-5, and $S\gtrsim5$ mJy sources at 850 \mum\ by an order
of magnitude or more. This problem is far from being unique to our
models, as we have discussed above. \citet{clements10} show that none
of the models that they compare with their SPIRE 250, 350 and 500 \mum\
count data agree with their results very well. Most of these models
are empirical ``backwards evolution'' models, but they also compare
with the semi-analytic model of \citet{lacey:08}, which predicts a
significant excess of luminous galaxies in all three SPIRE bands.

When comparing with estimates of luminosity functions at $z\sim
0.5$--2, we find deficits of luminous galaxies at high redshifts in
the rest 8 \mum\ and 24 \mum\ bands. In the mid-IR (8 and 24 \mum), it
is possible that there could be significant contamination from
obscured AGN, particularly at $z\sim 2$ \citep[e.g.][]{daddi:07}. The
agreement with the estimated total IR LF to $z\sim2$ is not terrible
(within the observational errors). It is important to remember that
these observational estimates rely on k-correcting from an observed
wavelength to the rest-frame in a messy part of the SED (particularly
in the case of 8 and 24 \mum), or on estimating a total IR luminosity
from a single or a limited number of observed wavelengths. These
conversions are themselves highly uncertain and in general rely on SED
templates. Therefore, it is encouraging that the agreement between our
models and the more directly observed quantity, the galaxy counts, is
in general superior to the agreement with the derived quantities
(rest-frame or total luminosity functions). In our companion
  paper (GSPD), we also show the redshift dependence of the build-up
  of the EBL at 24 $\mu$m, 70 $\mu$m, and 160 $\mu$m compared with
  available observations and find fairly good agreement.

By integrating over all galaxies, accounting for the redshifting and
dilution of light, we estimate the integrated Extragalactic Background
Light predicted by our models. As expected, our EBL predictions lie
close to the lower limits from integrated galaxy counts. The largest
uncertainties (model-to-model differences) in our predictions are in
the mid-IR, due to the limitations of available IR templates. These
will improve as more data from multi-wavelength observations are
synthesized. In particular, important constraints on this part of the
EBL, and correspondingly on the star formation history and dust SEDs
of galaxies, may be obtained from observations of GeV and TeV gamma
rays. High energy gamma rays are attenuated via electron-positron pair
production against the EBL. In principle, the cosmological history of
the EBL could be reconstructed by comparing observations of
high-energy sources at different redshifts to their known intrinsic
spectra. In a companion paper (GSPD), we provide a detailed analysis
of the implications of our predictions for current and future gamma
ray observations.

%===================================
\section*{Acknowledgments}
\begin{small}

We dedicate this paper to the memory of Donald Lee MacMinn. We warmly
thank James Bullock, Julien Devriendt, Bruno Guiderdoni, David Elbaz,
Fabio Fontanot for useful discussions. We thank Matthieu Bethermin and
Chris Kochanek for pointing out additional observational data to
include in our compilations, and we thank the anonymous referee for
comments that improved the paper. We thank Elysse Voyer and Timothy
Dolch for providing their observational data ahead of publication. RSS
and RCG thank UCSC for hospitality on numerous occasions during the
gestation of this paper.  RCG acknowledges support from a Fermi Guest
Investigator grant and a research fellowship from the SISSA
Astrophysics Sector.  JRP acknowledges support from NASA ATP grant
AST-1010033.

\end{small}
%=====================================

\bibliographystyle{mn} 
\bibliography{mn-jour,eblpaper1}

\end{document}